\documentclass[aps, prd, twocolumn, preprintnumbers, longbibliography, superscriptaddress, nofootinbib, floatfix,notitlepage]{revtex4-1}

\usepackage{amsmath}
\usepackage{bm}
\usepackage{amsfonts}
\usepackage{graphicx}
\usepackage{hyperref}
\usepackage[dvipsnames]{xcolor}
\usepackage[normalem]{ulem}
\usepackage{xspace}
\usepackage{relsize}
\usepackage{subeqnarray}
\usepackage{aas_macros}

\newcommand{\mycomment}[1]{}

\usepackage{appendix}
\usepackage{braket}

\newcommand{\primat}{\texttt{PRIMAT}\xspace}
\newcommand{\parthenope}{\texttt{PArthENoPE}\xspace}

\newcommand{\Neff}{N_\mathrm{eff}}

\newcommand{\Yp}{Y_\mathrm{p}}
\renewcommand{\DH}{\mathrm{D/H}}
\newcommand{\dd}{\mathrm{d}}
\newcommand{\obs}{\mathrm{obs}}
\newcommand{\flogE}{{\cal F}}
\newcommand{\bb}{{\cal B}}
\newcommand\transpose{T}
\newcommand{\bref}{\beta_\mathrm{ref}}
\newcommand{\kB}{k_{\rm B}}

\def\dd{{\rm d}} 

\hypersetup{
colorlinks=true,
linkcolor=blue,
linktoc=page,
citecolor=blue,
urlcolor=blue}

\hypersetup{colorlinks=true,
linkcolor=Mulberry,
citecolor=NavyBlue,
filecolor=Mulberry,
urlcolor=NavyBlue,
menucolor=BrickRed,
runcolor=Mulberry
}

\newcommand{\comment}[1]{}

\begin{document}

\title{Fifty shades of grayness: parametrizations of spectral distortions and applications in cosmology}

\newcommand{\SECONDAFF}{\affiliation{Departament de Física Teórica and IFIC, Universitat de València-CSIC, E-46100, Burjassot, Spain}}
\newcommand{\THIRDAFF}{\affiliation{Institut d’Astrophysique de Paris, CNRS UMR 7095, Sorbonne Universit\'e, 98 bis Bd Arago, 75014 Paris, France}}

\author{Gabriela Barenboim}
\email[Electronic address: ]{gabriela.barenboim@uv.es}
\SECONDAFF
\author{Julien Froustey}
\email[Electronic address: ]{julien.froustey@ific.uv.es}
\SECONDAFF
\author{Cyril Pitrou}
\email[Electronic address: ]{cyril.pitrou@iap.fr}
\THIRDAFF
\author{Héctor Sanchis}
\email[Electronic address: ]{hector.sanchis@uv.es}
\SECONDAFF

\date{\today}
\begin{abstract}
Thermal distribution functions can only be of the Fermi-Dirac or Bose-Einstein types, whereas distorted spectra encompass any possible deviations from these shapes. It is fruitful to devise parametrizations of these distortions with only a few parameters which depend on the physical system considered. A method proposed by Stebbins consists in describing a distorted spectrum as a sum of thermalized spectra with a distribution of temperatures, the moments of which are the parameters of interest. After revisiting and extending this approach by working at the level of the number density distribution instead of the standard spectrum, we build another method which consists in describing the distorted spectrum by a polynomial modulating a reference thermalized spectrum. The distortion parameters are then the coefficients of a decomposition on a suitable orthonormal polynomial basis. We advocate that the latter is computationally easier and allows to describe a wide range of distortions. With this formalism, we efficiently describe the standard distortions of the cosmological backgrounds of neutrinos and photons, and we obtain model-independent constraints on nonstandard distortions of these cosmological relics. 
\end{abstract}

\maketitle

\tableofcontents

\section{Introduction}

In cosmology, characterizing the spectral properties of fundamental radiation backgrounds is crucial for advancing our understanding of the Universe’s evolution and composition. The cosmic microwave background (CMB) is remarkably well described by a near-perfect blackbody spectrum~\cite{Fixsen:1996nj}, but tiny distortions of the $y$ and $\mu$ type are expected in the standard cosmological model due to the differential cooling of baryons~\cite{Hu:1995em,Chluba:2016bvg,Pitrou:2019hqg}. Furthermore, dissipation of acoustic waves would also induce distortions whose detection would provide information on the power spectrum of fluctuations at small scales~\cite{Chluba:2011hw}. More generally, any additional energy injection or depletion in the early Universe plasma would also give rise to additional $y$-type and $\mu$-type distortions, hence providing information about physics prior to recombination~\cite{Chluba:2013vsa,Chluba:2013wsa}. The cosmic neutrino background (C$\nu$B) also exhibits percent-level distortions~\cite{Mangano:2005cc, Grohs:2015tfy, Akita:2020szl,Froustey:2020mcq, Bennett:2020zkv} generated around their decoupling, and emissions from astrophysical sources like interstellar dust exhibit even more complex spectral features. 

Equilibrium solutions of a particle species are related to its fundamental statistical properties, and using the grand canonical ensemble, it can be shown that it is either of the Fermi-Dirac type or of the Bose-Einstein type. This is also reflected by the structure of any collision term in a Boltzmann equation. Indeed, if we consider four species with respective isotropic distribution functions $f_i(E_i)$ with $i=1,\dots,4$, then the collision term describing reactions of the type $1+2 \leftrightarrow 3+4$ is the sum of forward and backward rates with factors taking into account Pauli blocking or Bose enhancement. Detailed balance of the forward and backward rates implies the equilibrium condition
\begin{equation}
\frac{f_1(E_1)}{[1\mp f_1(E_1)]}\frac{f_2(E_2)}{[1\mp f_2(E_2)]} = \frac{f_3(E_3)}{[1\mp f_3(E_3)]}\frac{f_4(E_4)}{[1\mp f_4(E_4)]}\,,
\end{equation}
where throughout this article the upper sign is for fermions and the lower sign is for bosons. With the conservation of energy $E_1+E_2 = E_3+E_4$, we check that any of the factors appearing must be of the type $\exp[-(E_i-\mu_i)/\kB T]$ with $\mu_1+\mu_2 = \mu_3+\mu_4$. Therefore equilibrium distributions are necessarily of the type 
\begin{equation}\label{DefF}
f(E) = \bb\left(\frac{E-\mu}{\kB T}\right)\,,\qquad \bb(x) \equiv \frac{1}{{\rm e}^{x} \pm 1}\,.
\end{equation}
Throughout this article, we call {\it blackbody} any such spectrum with vanishing chemical potential, even for fermions. Any deviation from the thermodynamic equilibrium contains information about out-of-equilibrium physics, the chemical potential being constrained by the types of reactions which are possible.

Distortions can generically be described by a frequency-dependent emissivity factor which multiplies the blackbody spectrum. In general, if the emissivity factor is constant across all frequencies, the spectrum is gray~\cite{Barenboim:2025vrc} and of the form 
\begin{equation}\label{DefGrayBody}
f_{\rm gray}(E) = (1-g) \, \bb\left(\frac{E}{\kB T}\right)\,,
\end{equation}
where $g$ is the constant graybody parameter. However, the frequency dependence is \emph{a priori} general, hence the vector space of distortions is infinite dimensional. One would nonetheless like to capture the essential features of the distortion with just a few numbers, especially if the distortion is small as in the cases mentioned above.

A proposal~\cite{Stebbins:2007ve} consists in describing a spectrum as a sum of equilibrium spectra. In principle we could consider a two-dimensional sum of $f^{\rm eq}_{T,\mu}(E)$ with a continuous weight function $p(T,\mu)$. It is nonetheless simpler to describe a distorted spectrum with $p(T) \equiv p(T,0)$ only. This temperature distribution is called the \emph{temperature transform} (TT) and the distortion of the spectrum is described by the moments of $p(T)$. In practice one rather considers the distribution of $q(\ln T) = T p(T)$ and its moments in the variable $\ln T$ to take advantage of the fact that gravitational effects shift $\ln E$ by the same amount~\cite{Pitrou:2014ota}. A major drawback of this TT method is that obtaining the temperature transform from a given spectrum is not straightforward. It was initially proposed in~\cite{Stebbins:2007ve} to compute directly the TT moments from a spectrum, and after summarizing this method we provide in Sec.~\ref{SecTT} a different technique based on Fourier transforms, which allows us to evaluate the temperature transform itself directly, and then compute from it the various moments.

In Sec.~\ref{SecPE}, we present an alternative description of distortions using an expansion on polynomials, which are orthonormal with respect to a scalar product based on the blackbody number density. 
We gather details on the construction of the polynomial basis in Appendices~\ref{ap:PolynomialCoefficients} and \ref{app:recursion}. We then compare in Sec.~\ref{SecIdeal} the TT and OPE descriptions on idealized distortions, and in particular in the case of $y$-type and $\mu$-type distortions. This allows to show that only a few coefficients of the orthonormal polynomial expansion (OPE) are needed to accurately describe most spectral distortions. 

Finally, we apply our formalism to two examples in the cosmological context in Sec.~\ref{SecApplications}. We decompose the distortion of the cosmological neutrino background generated at the epoch of their decoupling and we obtain constraints on the coefficients of the OPE from the cosmic microwave background (CMB) anisotropies and big bang nucleosynthesis (BBN) data. We also decompose the spectral distortions of the CMB monopole and constrain the coefficients from FIRAS data~\cite{Fixsen:1996nj}. When constraining energy injection in the early Universe, we also show that the OPE provides a good description of the redshift dependent Green transfer functions which describe the resulting CMB monopole spectrum. In Sec.~\ref{Massive_Case}, we discuss how to generalize the OPE method to the case of massive particles. Finally, we summarize and conclude in Sec.~\ref{Conclusion}.

\section{Temperature transforms}\label{SecTT}

We generally call ``temperature transforms" the representation of a given spectrum as a combination of thermal spectra. The distribution in temperature space allows one to compute a set of moments, which one can use to describe the distorted spectrum. Specifically, keeping only a limited number of moments allows one to use only a finite number of parameters to describe the distortion. There are however several choices of transforms that we review in this section.

\subsection{Standard temperature transform (TT)}

The simplest temperature transform consists in describing a given spectrum $f(E)$ as the sum of blackbody spectra. It is therefore defined as
\begin{equation}
f(E) = \int_0^\infty p(T) \, \bb\left(\frac{E}{\kB T}\right) \dd T\,,
\end{equation}
and the moments of the TT are defined by
\begin{equation}
p_n \equiv \int_0^\infty  p(T) \, T^n \, \dd T\,.
\end{equation}
The energy moments of the spectrum, denoted $f_n$, are directly related to the TT moments since we obtain
\begin{equation}\label{Deffn}
f_n \equiv \int_0^\infty E^n f(E) \dd E = p_{n+1} I_n\,,
\end{equation}
when using the blackbody moments (defined for $n>0$ for bosons and $n>-1$ for fermions)
\begin{align}\label{DefIn}
I_n &\equiv \int_0^\infty x^n \bb(x) \dd x\\
&=\Gamma(n+1)\begin{cases}
\zeta(n+1)  &\ \text{[bosons]}\,,\\
\eta(n+1)  &\ \text{[fermions]}\,,\nonumber
\end{cases}
\end{align}
where $\Gamma(n)$, $\zeta(n)$ and $\eta(n)$ are respectively the gamma, Riemann zeta, and Dirichlet eta functions, the latter two being related by $\eta(n+1) = \zeta(n+1)(1-2^{-n})$.
Their analytic expressions are related to the more general Fermi-Dirac or Bose-Einstein integrals 
\begin{equation}\label{DefJn}
J_n(\xi) \equiv\frac{1}{\Gamma(n+1)} \int_0^\infty x^n \bb(x-\xi) \dd x =\mp{\rm Li}_{n+1}(\mp {\rm e}^\xi)\,,
\end{equation}
where ${\rm Li}$ is the polylogarithm function, via $I_n \equiv \Gamma(n+1) J_n(0)$. We recall that in \eqref{DefJn}, the upper (lower) sign is for fermions (bosons).

Since we will consider cosmological applications in Sec.~\ref{SecApplications}, it is noteworthy to discuss the effect of redshifting on the spectrum and its temperature transform. When energies are redshifted as 
\begin{equation}
\label{eq:redshift}
E \to E_z=E/(1+z)\,,
\end{equation}
then $f(E) \to f_z(E)=f(E(1+z))$ hence $ p(T) \to p_z(T)=(1+z)p(T(1+z))$, and the moments transform as 
\begin{equation}
f_n \to \frac{f_n}{(1+z)^{n+1}}\,,\quad p_n \to \frac{p_n}{(1+z)^n}\,.\label{Transfofn}
\end{equation}

\subsection{Log-temperature transform (LTT)}\label{SecLTT}
 
We now review the log-temperature transform, introduced in~\cite{Stebbins:2007ve}, which has simpler transformation properties under redshifting, and enables, as we will see later, the use of Fourier transforms to simplify calculations.

\subsubsection{Definitions and moments}

The log-temperature transform is the convolution
\begin{equation}
\label{eq:fnu_superposition}
    f(E) = \int_{-\infty}^{+\infty}{q(\mathcal{T}) \, \bb\left({\rm e}^{{\cal E}-{\cal T}}\right) \mathrm{d}{\cal T}} \,,
\end{equation}
where we use the log variables
\begin{equation}
\label{eq:logvariables}
{\cal E} = \ln (E/\kB T_0)\,,\quad {\cal T} = \ln (T/T_0)\,,
\end{equation}
for some arbitrary reference temperature $T_0$. It is related to the TT via $T p(T) = q({\cal T})$. Note that a change in $T_0$ corresponds to a shift of $q(\mathcal{T})$ along the $\mathcal{T}$ axis.

The LTT states how to obtain a spectrum given a log-temperature transform $q(\mathcal{T})$. However the inverse relation is not straightforward, especially because the set of $\bb\left({\rm e}^{{\cal E}-{\cal T}}\right) = \bb(E {\rm e}^{-{\cal T}}/\kB T_0)$ indexed by the variable $\mathcal{T}$ is not a basis.

Let us define the moments of the LTT which characterize the distortion by
\begin{equation}
q_n\equiv \int_{-\infty}^{+\infty} {q}({\cal T}) {\cal T}^n\dd {\cal T}\,. \label{def_q_moments}
\end{equation}
First, the LTT might not be suitably normalized and we are led to consider the graybody parameter $g$~\cite{Stebbins:2007ve,Barenboim:2025vrc}
\begin{equation}
(1-g) \equiv q_0 = \int_{-\infty}^{+ \infty}{q(\mathcal{T})  \dd{\mathcal{T}}}\,,
\end{equation}
such that $q({\cal T})/(1-g)$ can be considered as a properly normalized log-temperature distribution. The average log-temperature is defined by
\begin{equation}
\overline{\mathcal{T}} = \frac{q_1}{q_0}=\int_{-\infty}^{+ \infty} \frac{q(\mathcal{T})}{1-g} \mathcal{T} \dd{\mathcal{T}}\,.
\end{equation}
It can subsequently be used as a reference temperature to define the central moments $u_n$ as
\begin{equation}
    u_n \equiv \langle (\mathcal{T} - \overline{\mathcal{T}})^n \rangle = \int_{-\infty}^{+ \infty} \frac{q(\mathcal{T})}{1-g} \, (\mathcal{T} - \overline{\mathcal{T}})^n \dd{\mathcal{T}} \, ,
\end{equation}
with $u_0=1$ and $u_1=0$ by construction.

When energies are redshifted as in~\eqref{eq:redshift}, that is when $\mathcal{E} \to \mathcal{E}_z =\mathcal{E} - \ln (1+z)$, then 
\begin{equation}
q({\cal T}) \to q_z({\cal T}) = q[{\cal T}+\ln (1+z)]\,,
\end{equation}
and the overall shape of the LTT is preserved. This means that all central moments and the graybody parameter are invariant under redshifting, as it only shifts the average energy according to $\overline{\mathcal{T}} \to \overline{\mathcal{T}}-\ln(1+z)$. For the same reason, the grayness $g$ and the central moments $u_n$ are independent from the choice of $T_0$.

With $\bar{T}/T_0 \equiv \exp \overline{\cal T}$, the moments of the LTT are related to the moments of the TT thanks to~\cite{Stebbins:2007ve}
\begin{equation}
p_n = \bar{T}^n \, (1-g) \, \sum_m n^m \frac{u_m}{m!} =T_0^n \, \sum_m n^m \frac{q_m}{m!}\,.
\end{equation}
The inverse relation reads
\begin{subequations}
\begin{align}
u_n &= n!\sum_{m=0}^\infty \frac{s(m,n)}{m!} \frac{p^{\bar{T}}_m}{\bar{T}^m}\,,\\
q_n&= n!\sum_{m=0}^\infty \frac{s(m,n)}{m!} \frac{p^{T_0}_m}{T_0^m}\,,
\end{align}
\end{subequations}
where the $s(m,n)$ are Stirling numbers of the first kind, and with moments centered around a given temperature $T_c$ defined by
\begin{equation}
p^{T_c}_n\equiv\langle(T-T_c)^n\rangle = \sum_{m=0}^n {n\choose m}  (-T_c)^{n-m}p_m\,.
\end{equation}

\subsubsection{Spectrum reconstruction}

The log-temperature transform can be reconstructed from its moments $q_n$ or from its central moments $u_n$ (along with $\overline{\cal T}$ and $g$) thanks to
\begin{subequations}
\begin{align}\label{qfromun}
q(\mathcal{T}) &= (1-g)\sum_{n=0}^{\infty}{\frac{(-1)^n}{n!} \, u_n \, \delta^{(n)}(\mathcal{T}-\bar{\mathcal{T}})}\,,\\
&=\sum_{n=0}^{\infty}{\frac{(-1)^n}{n!} \, q_n \,\delta^{(n)}(\mathcal{T})}\,.
\end{align}
\end{subequations}
With integration by parts, the initial spectrum is then obtained from the moments via
\begin{subequations}
\label{eq:fnu_from_moments}
\begin{align}
   f(E) &= (1-g)\sum_{n=0}^{\infty}\frac{u_n}{n!}\, \left.\frac{\dd^n}{\dd \mathcal{T}^n} \bb(\beta E) \right|_{{\cal T} = \overline{\cal T}} \,,\label{eq:fnu_from_moments1}\\
   &=\sum_{n=0}^{\infty}\frac{q_n}{n!}\left.\frac{\dd^n}{\dd \mathcal{T}^n} \bb(\beta E) \right|_{{\cal T} = 0}\,.\label{eq:fnu_from_moments2}
\end{align}
\end{subequations}
We introduced, for brevity, the notation $\beta \equiv 1/\kB T$. For the arbitrary reference temperature $T_0$, we will also use $\beta_0 \equiv 1/\kB T_0$ such that $\beta$ depends on $\mathcal{T}$ as $\beta = \beta_0e^{-\mathcal{T}}$.
A practical description of a distorted spectrum consists in keeping only a few multipoles up to $n_{\rm max}$. Note that the order-$n$ truncation of the reconstruction from the central moments is independent from the choice of $T_0$ (assuming the central moments are known without error), whereas the reconstruction from the non-centered moments does not have this property.

\subsubsection{LTT moment extraction à la Stebbins}

The temperature transform is a convolution since
\begin{equation}
    \flogE({\cal E})\equiv f(E) = \int_{-\infty}^{+\infty} k({\cal E}-{\cal T}) q({\cal T}) \dd {\cal T}\,,
\end{equation}
with the kernel
\begin{equation}\label{oldkernel}
k(z) \equiv \left(e^{e^z} \pm 1 \right)^{-1}\,.
\end{equation}
The moments of the functions, defined by
\begin{subequations}
\begin{align}
\flogE_n &\equiv \int_{-\infty}^{+\infty} \flogE({\cal E}){\cal E}^n \dd {\cal E}\,,\\
 k_n &\equiv \int_{-\infty}^{+\infty} k(z) z^n\dd z\,, 
\end{align}
\end{subequations}
are  related via
\begin{equation}
\flogE_n = \sum_{p=0}^n \binom{n}{p} k_{n-p} q_{p}\,.
\end{equation}
In practice, the $q_n$ (which are the moments of the log-temperature distribution) should be obtained iteratively via
\begin{equation}\label{qnfromFn}
q_n = \frac{\flogE_n-\sum_{p=0}^{n-1} \binom{n}{p} k_{n-p} q_p}{k_0}
\end{equation}
once the moments $\flogE_n$ (which are directly moments of the spectrum) are known. The $\overline{\cal T}$, $g$ and $u_n$ moments (for $n\geq 2$)  are then obtained via
\begin{subequations}\label{Relqu}
\begin{align}
g&= 1- q_0\,,\quad \overline{\cal T} = q_1/q_0\,,\label{Relqua}\\
(1-g)u_n &= \sum_{m=0}^n {n\choose m} (- \overline{\cal T})^{n-m} q_m \,.\label{Relqub}
\end{align}
\end{subequations}

Unfortunately, the kernel $k(z)$ does not vanish when $E \to 0$ (or ${\cal E}\to -\infty$), such that the moments are not well defined and the above procedure cannot always be used in practice. It was proposed in~\cite{Stebbins:2007ve} to use a regularization method, so as to get the moments directly from the spectrum, without computing the temperature transform. We will use this approach on an example in Sec.~\ref{SecIdeal} and show its limitations. In the next subsection, we improve on this method by using a slightly different kernel, which has a proper behavior as $E \to 0$.

\subsection{Number density log-temperature transform (NLTT)}\label{SecNLTT}

Fundamentally, the problem encountered with the kernel $k(z)$ when $E \to 0$ arises because the temperature transform is defined from the distribution function $f(E)$, instead of the number density spectral function. We now review how the log-temperature transform can be modified to be defined with the number density, and how this approach avoids the problems associated with $k(z)$.

\subsubsection{General formalism}

Let us consider the number density log-temperature transform (NLTT) defined by 
\begin{equation}
\begin{aligned}
    \tilde{\flogE}({\cal E})\equiv {\rm e}^{2{\cal E}} \flogE({\cal E}) &= \left(\beta_0 E\right)^2f(E) \\
    &= \int_{-\infty}^{+\infty} \tilde{k}({\cal E}-{\cal T}) \tilde{q}({\cal T}) \dd {\cal T} \, ,
    \label{logT_convolution_tilde}
\end{aligned}
\end{equation}
where the new kernel and the modified temperature transform are
\begin{equation}
    \tilde{k}(z) = e^{2z} \left(e^{e^z} \pm 1 \right)^{-1}\,,\qquad \tilde{q}({\cal T}) \equiv {\rm e}^{2{\cal T}} q({\cal T})\,.
    \label{newkernel}
\end{equation}
We call it the number density log-temperature transform because the number density distribution $n(E)$ fulfills $n(E) \propto E^2 f(E)$ for massless or ultrarelativistic particles.\footnote{For massive particles, this can no longer be interpreted as a number density, but the method is in principle still applicable.} For $E \to 0$, $E^2 f(E)$ goes to zero, or equivalently $\lim_{z \to - \infty} \tilde{k}(z) = 0$ sufficiently fast so that the moments of this new kernel are well defined.

We then work at the level of the Fourier transforms, since a convolution becomes a product in Fourier space. We need only to perform a division of functions to obtain the Fourier transform of $\tilde{q}$ from that of $\tilde{\flogE}$. When dividing the Fourier transforms, division by zero issues appear, since the Fourier transform of the kernel tends to vanish for $z \rightarrow \pm \infty$. This numerical problem can be avoided by regularizing the division, as is done in Wiener deconvolution. If $F$ (resp.~$K$ and $Q$) is the Fourier transform of $\tilde{\flogE}$ (resp.~$\tilde{k}$ and $\tilde{q}$), one can write (for all points where $K \neq 0$) that $Q=F/K = FK^*/|K|^2$. Then, to regularize this expression in the cases where $K = 0$, one introduces the (small) regularization parameter $\alpha$ and the replacement 
\begin{equation}
    Q=\frac{F}{K}=\frac{FK^*}{|K|^2} \longrightarrow \frac{FK^*}{|K|^2 + \alpha} \, .
    \label{Wiener_regularization}
\end{equation}
The NLTT $\tilde{q}(\mathcal{T})$ can then be obtained by applying the inverse Fourier transform on $Q$.

In principle, once we have the modified log-temperature transform $\tilde{q}(\mathcal{T})$, we can simply use Eq.~\eqref{newkernel} to calculate the original temperature transform $q(\mathcal{T})$. In practice, we find that the exponential in the relation between them tends to amplify numerical errors at large negative values of $\mathcal{T}$, making $q(\mathcal{T})$ unusable for the purpose of calculating moments, so it is generally better to work with $\tilde{q}(\mathcal{T})$.

\subsubsection{Moments}

We can define moments of the NLTT as we did for the LTT. The modified moments, central moments, average temperature and grayness are simply  defined as 
\begin{subequations}
\begin{align}
\tilde{q}_n &\equiv \int_{-\infty}^\infty \mathrm{d}\mathcal{T} \mathcal{T}^n \tilde{q}(\mathcal{T})\,,\label{etatilde_moments_def}\\
\widetilde{\overline{\mathcal{T}}} &\equiv \tilde{q}_1/\tilde{q}_0\,,\\
\tilde{u}_n &\equiv \int_{-\infty}^{+ \infty} \frac{\tilde{q}(\mathcal{T})}{\tilde{q}_0} \, (\mathcal{T} - \widetilde{\overline{\mathcal{T}}})^n \dd{\mathcal{T}} \,,\\
(1-\tilde{g}) &\equiv {\rm e}^{-2\widetilde{\overline{\mathcal{T}}}}\tilde{q}_0\,,\label{Defgtildeadhoc}
\end{align}
\end{subequations}
with $\tilde{u}_0=1$ and $\tilde{u}_1=0$ by construction. These definitions ensure that for a pure graybody spectrum, $\tilde{g} = g$ and $\widetilde{\overline{\mathcal{T}}}=\overline{\mathcal{T}}$.

Under redshifting, the NLTT transforms as
\begin{equation}
\tilde{q}({\cal T}) \to \tilde{q}_z({\cal T}) = 
(1+z)^{-2} \tilde{q}[{\cal T}+\ln (1+z)]
\end{equation}
hence all central moments $\tilde{u}_n$ are invariant. The graybody parameter $\tilde{g}$ is also invariant thanks to the definition~\eqref{Defgtildeadhoc}. Therefore, under redshifting the transformation reduces to shifting the average energy as  $\widetilde{\overline{\mathcal{T}}} \to \widetilde{\overline{\mathcal{T}}}-\ln(1+z)$. Again, this also implies that $\tilde{g}$ and $\tilde{u}_n$ do not depend on the arbitrary $T_0$ introduced in \eqref{eq:logvariables}.

The relation between moments and central moments is similar to~\eqref{Relqub}, that is
\begin{equation}
\tilde{u}_n = \sum_{m=0}^n {n\choose m} (- \widetilde{\overline{\cal T}})^{n-m} \frac{\tilde{q}_m}{\tilde{q}_0} \,.\label{centeredtildefromqtilde}
\end{equation}
The reconstruction of the distribution in terms of the central moments is finally similar to~\eqref{eq:fnu_from_moments}, its truncations are also independent from $T_0$ and it reads\footnote{This reconstruction is not affected by the exponential amplification of numerical noise mentioned after Eq.~\eqref{Wiener_regularization}, because the factor $\exp(- 2 \mathcal{T})$ is evaluated here for ${\cal T} = \widetilde{\overline{\cal T}}$, which is typically not a large negative number.} 
\begin{equation}\label{RecconstructionFromTilde}
f(E)=  (1-\tilde{g}) e^{2\widetilde{\overline{\cal T}}}\sum_{n=0}^{\infty}{\frac{\tilde{u}_n}{n!} \frac{\dd^n}{\dd {\cal T}^n} \Big[e^{-2{\cal T}} \, \bb(\beta E)\Big]_{{\cal T} = \widetilde{\overline{\cal T}}}}\, .
\end{equation}

The moments of $q(\mathcal{T})$ and $\tilde{q}({\cal T})$ are related. Indeed, from~\eqref{newkernel}, and performing a Maclaurin expansion of the exponential, we get
\begin{equation}
    q_n = \sum_{k=0}^\infty \frac{(-2)^k}{k!} \tilde{q}_{n+k}\,,\qquad \tilde{q}_n = \sum_{k=0}^\infty \frac{2^k}{k!} {q}_{n+k}
    \label{moment_relation_series}\,.
\end{equation}

\subsubsection{Moment-generating function}\label{SecUnMoments}

The main advantage of the aforementioned method is that it allows to calculate the temperature transform itself, from which we can subsequently extract the moments which describe the distortion. 

However, even though we can calculate $\tilde{q}({\cal T})$, getting the moments directly from the definition \eqref{etatilde_moments_def} turns out to be numerically problematic, especially at higher orders, where the integral gives a lot of weight to large values of $\mathcal{T}$, where $\tilde{q}({\cal T})$ is small and can be dominated by numerical noise. Instead, we find that it is better to calculate the moments via the imaginary moment-generating function, defined as 
\begin{equation}
    M_{\tilde{q}}(t) \equiv \int_{-\infty}^{+\infty} e^{it\mathcal{T}} \tilde{q}(\mathcal{T}) \mathrm{d}\mathcal{T}\,.
    \label{moment_generating_function_imag}
\end{equation}

The moments can be computed from the derivative of this function according to
\begin{equation}
    \tilde{q}_n = i^{-n} \frac{\mathrm{d}^n M_{\tilde{q}}}{\mathrm{d}t^n}(t=0)\,.
    \label{moments_from_genfunction}
\end{equation}
Central moments are then obtained thanks to~\eqref{centeredtildefromqtilde} and the spectrum reconstruction is performed with~\eqref{RecconstructionFromTilde}.

The spectrum can also be reconstructed from the moments of $q({\cal T})$ with~\eqref{Relqu} and~\eqref{eq:fnu_from_moments}, since they are related to the moments of $\tilde{q}({\cal T})$ via~\eqref{moment_relation_series}. However, in practice we can only calculate a finite amount of moments, so using~\eqref{moment_relation_series} requires a truncation. 
This alternative method to compute the moments of the LTT for a given spectrum is illustrated in Sec.~\ref{SecIdeal} on idealized cases. 

\section{Polynomial description}\label{SecPE}

A more direct description of a distorted spectrum consists in expanding it onto a suitably normalized polynomial basis, and keeping only a few coefficients.

\subsection{Orthonormal basis}

For reasons which become clear hereafter, we consider a set of polynomials which are orthonormal for the Fermi-Dirac or Bose-Einstein weights. The scalar product with such weights is defined by
\begin{equation}
    \Braket{f | g} = \int_0^\infty w(x) f(x) g(x) \mathrm{d}x\,,\quad w(x)\equiv \frac{x^2}{e^x \pm 1}\,.
    \label{ScalarProduct_def}
\end{equation}
These weights are not traditionally used, such that the associated orthonormal polynomials do not belong to a known class of special functions. We do not explore here their mathematical properties, such as whether they are solutions of a specific differential equation, whether their generating function has a closed form, etc.

\begin{figure}[!ht]
    \centering
    \includegraphics[width=\columnwidth]{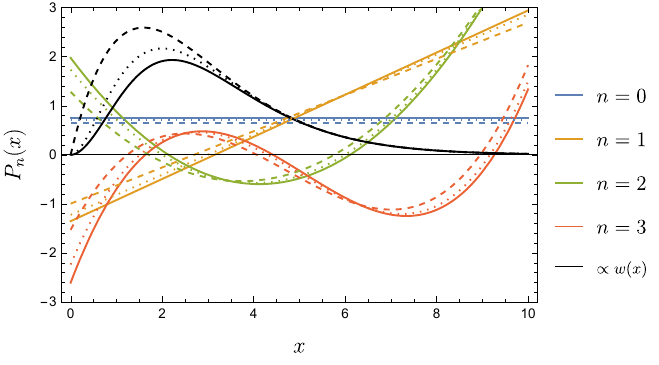}
    \caption{Orthonormal polynomials with respect to the weights~\eqref{ScalarProduct_def} with plus sign for fermions (continuous lines) and minus sign for bosons (dashed lines). The dotted lines are the normalized Laguerre polynomials~\eqref{Laguerre}. The black lines are the corresponding weight functions (rescaled for better visibility).}
    \label{FigPn}
\end{figure}

We define $P_i(x)=\sum_{p=0}^i A_{ip} x^p$ which are orthonormal with respect to this scalar product, that is, which satisfy
\begin{equation}
    \Braket{P_i | P_j}= \sum_{p=0}^i\sum_{q=0}^j A_{ip} A_{jq} \langle x^p|x^q\rangle =\delta_{ij}\,.
    \label{Pi_orthogonality}
\end{equation} 
Using a vector notation ${\bm P} = (P_0,P_1,P_2,\dots)^T$, ${\bm X}=(x^0,x^1,x^2,\dots)^T$, the notation ${\bm A}$ for the lower triangular matrix whose components are $A_{ip}$ and the notation ${\bm H}$ for the Hankel matrix whose components are expressed with integrals~\eqref{DefIn} as
\begin{equation}\label{HankelValues}
H_{pq} \equiv \langle x^p|x^q \rangle=I_{p+q+2}\,,
\end{equation}
this is rephrased as 
\begin{equation}\label{DefPiPj}
{\bm P} = {\bm A} \cdot {\bm X}\,,\qquad {\bm A} \cdot {\bm H}\cdot {\bm A}^{\transpose} = \mathbb{I}\,.
\end{equation}
The components $A_{ip}$ can be found by Gram-Schmidt orthogonalization, and we report their numerical values in Appendix~\ref{ap:PolynomialCoefficients}. An alternative procedure consists in using the recursion relation satisfied by the polynomials, as detailed in Appendix~\ref{app:recursion}. The components of the inverse matrix ${\bm A}^{-1}$, which is also lower triangular, can be obtained by the forward substitution method, and from~\eqref{DefPiPj} they also satisfy ${\bm A}^{-1} = {\bm H}\cdot {\bm A}^\transpose$ which implies in particular
\begin{equation}\label{PureMagic}
(A^{-1})_{p0} = H_{p0}A_{00} = I_{p+2}A_{00}\,.
\end{equation}
The first polynomials are depicted in Fig.~\ref{FigPn}.

Let us introduce a reference temperature $T_\mathrm{ref}$, which can in principle be freely chosen, and define 
\begin{equation}
\beta_\mathrm{ref}\equiv\frac{1}{\kB T_\mathrm{ref}}\,.
\end{equation}
For any $T_\mathrm{ref}$, the $P_i$ satisfy
\begin{equation}
    \int_0^\infty \frac{E^2}{e^{\beta_\mathrm{ref}E} \pm 1} P_i\left(\beta_\mathrm{ref}E\right) P_j\left(\beta_\mathrm{ref}E\right)\mathrm{d}E =  \beta_\mathrm{ref}^{-3} \delta_{ij}\,.
    \label{P_x_over_T_orthogonality}
\end{equation}
We can interpret this as a $T_\mathrm{ref}$-dependent scalar product, with respect to which the polynomials $\beta_\mathrm{ref}^{3/2} P_i(\beta_\mathrm{ref} E)$ form an orthonormal basis. 

Since the FD or BE weight functions~\eqref{ScalarProduct_def} do not differ much from the weights $w(x)=x^2 \exp(-x)$, the polynomials do not differ much from the normalized associated Laguerre polynomials 
\begin{equation}\label{Laguerre}
Q^{(2)}_n(x) \equiv  \frac{(-1)^n}{\sqrt{(n+1)(n+2)}} L^{(2)}_n(x)\,,
\end{equation}
which are associated with this weight, as can be seen on Fig.~\ref{FigPn}.

\subsection{Expansion on polynomial basis}\label{SecOPE}

We then define the decomposition of the spectrum $f(E)$, with an associated number density distribution $n(E)$ and a reference temperature $T_\mathrm{ref}$, as 
\begin{equation}
\begin{aligned}
    E^2f(E) &= \frac{2\pi^2(\hbar c)^3}{g_s} n(E) \\
    &=\frac{E^2}{e^{\beta_\mathrm{ref}E} \pm 1} \sum_{i=0}^\infty c_i P_i\left(\beta_\mathrm{ref}E\right)\,,
\end{aligned}
    \label{Spectrum_decomposition}
\end{equation}
where $g_s$ is the number of spin degrees of freedom. The prefactor in front of $n(E)$ is chosen such that the number density $n$ is given below by Eq.~\eqref{eq:densities}. The coefficients $c_i$ can be calculated by using property \eqref{P_x_over_T_orthogonality}, which yields 
\begin{align}
    c_i &= \beta_\mathrm{ref}^{3} \int_0^\infty E^2 f(E) P_i\left(\beta_\mathrm{ref}E\right)\mathrm{d}E\nonumber\\
    &= \sum_{p=0}^i A_{ip} \beta_\mathrm{ref}^{3+p}f_{p+2}\,,
    \label{coeffs_calc}
\end{align}
where the energy moments are defined in~\eqref{Deffn}, hence showing that the OPE is equivalent to linear combinations of the TT moments. Any truncation of the decomposition~\eqref{Spectrum_decomposition} provides an approximation for the full spectrum.

\subsection{Properties}

The total number and energy densities associated with a given spectrum are 
\begin{equation}
\label{eq:densities}
n \equiv \int_0^\infty n(E) \dd E\,,\qquad \rho \equiv \int_0^\infty E n(E) \dd E\,.
\end{equation}
Since any monomial $E^n$ can be decomposed as a sum on the $P_i(\beta_{\rm ref} E)$ with $0 \leq i \leq n$, and given the orthonormality property~\eqref{Pi_orthogonality}, we obtain
\begin{subequations}
\label{eq:nandrhofromci}
\begin{align}
n &= \frac{g_s (\kB T_{\rm ref})^3}{2 \pi^2 (\hbar c)^3} \, \frac{c_0}{A_{00}}\,,\label{nfromc0}\\
\rho &= \frac{g_s (\kB T_{\rm ref})^4}{2 \pi^2 (\hbar c)^3}\left[\frac{c_1}{A_{11}}+c_0\frac{(-A_{10})}{A_{11}A_{00}}\right]\,,\label{rhofromc01}
\end{align}
\end{subequations}
where we used $1 = A_{00}^{-1}P_0(x)$ and $x = A_{11}^{-1}[P_1(x)-A_{10}/A_{00}P_0(x)]$. 

A key property of the decomposition~\eqref{Spectrum_decomposition}, which stems from the choice of the weight~\eqref{ScalarProduct_def}, is that the number density is solely determined by $c_0$, and only $c_0$ and $c_1$ contribute to the energy density (for a given reference temperature). This is the key property that allows us to describe distorted spectra with only a few coefficients, as we show in the following. We can generalize this to any energy moment, since~\eqref{coeffs_calc} is recast as
\begin{equation}\label{MagicFormula}
f_{p+2}=\int_0^\infty E^{p+2} f(E) \dd E = (\kB T_{\rm ref})^{3+p} \sum_{i=0}^p (A^{-1})_{pi} c_i\,.
\end{equation}
When applied to a blackbody spectrum, this relation implies the identity~\eqref{PureMagic}.

\subsection{Reference temperature}
\label{subsec_Tref}

As mentioned earlier, the reference temperature $T_\mathrm{ref}$ can be chosen freely, and there is no unique perfect choice. Given a spectrum, one often needs to tune it to whatever yields the best convergence. Sets of coefficients with different reference temperatures $T_a$ and $T_b$ are related via 
\begin{equation}\label{MagicFlow1}
\left.c_i\right|_{T_b}  = \sum_{p=0}^i \sum_{j=0}^p A_{ip} \left(\frac{\beta_b}{\beta_a}\right)^{3+p} (A^{-1})_{pj} \left.c_j\right|_{T_a} \, ,
\end{equation}
as is directly seen from~\eqref{coeffs_calc} and~\eqref{MagicFormula}, when using that the energy moments $f_p$ do not depend on the reference temperature chosen. In matrix notation this reads
\begin{equation}\label{MagicFlow2}
\left.{\bm C}\right|_{T_b} = {\bm A}\cdot\exp[-{\bm D} \ln(T_b/T_a)] \cdot {\bm A}^{-1}\cdot\left.{\bm C}\right|_{T_a} \, ,
\end{equation}
where the diagonal matrix ${\bm D}$ has components $D_{pq}=(3+p)\delta_{pq}$. In addition, using~\eqref{Transfofn} in~\eqref{coeffs_calc}, it is immediate to see that under redshifting the $c_i$ transform exactly like when $T_{\rm ref} \to T_{\rm ref}(1+z)$, that is the redshifted coefficients are related to the initial ones via
\begin{equation}\label{MagicRedshift}
{\bm C}_z = {\bm A}\cdot\exp[-{\bm D} \ln(1+z)] \cdot {\bm A}^{-1}\cdot{\bm C}\,.
\end{equation}
Equivalently, the $c_i$ are invariant under redshifting provided the reference temperature is rescaled as $T_{\rm ref} \to T_{\rm ref}/(1+z)$ at the same time, as can be seen by combining~\eqref{MagicRedshift} with~\eqref{MagicFlow2}.

In this subsection, we present three possibilities for the choice of $T_\mathrm{ref}$. For a given number and energy density, there is an infinity of parameters $\{c_0, c_1, T_\mathrm{ref}\}$ that satisfy Eqs.~\eqref{eq:nandrhofromci}. The three possibilities we present in the following consist in making a simple additional choice, which then determines uniquely the values of $\{c_0, c_1, T_\mathrm{ref}\}$ given $n$ and $\rho$.
With any of these choices, redshifting leads to the same set of $c_i$ but with a redshifted $T_\mathrm{ref}$.

\subsubsection{Number density choice}

We can decide that the reference temperature is the one of the blackbody which has the same number density as the full spectrum. By construction this corresponds to $c_0 =1/A_{00}$ such that 
\begin{equation}
f(E) = \frac{1}{{\rm e}^{\bref E}\pm 1}\left[1 + \sum_{i=1}^\infty c_i P_i(\bref E)\right] \, ,
\end{equation}
Note that, for this choice, the zeroth order truncation of the expansion is a blackbody spectrum.
For any other choice, the zeroth order truncation will be a rescaled blackbody spectrum, i.e., a graybody spectrum.

In practice, from a known spectrum and its associated number density $n$, the reference temperature corresponding to this ``number density choice" is read from Eq.~\eqref{nfromc0},
\begin{equation}\label{Trefnrelation}
(\kB T_{\rm ref})^3 = \frac{2\pi^2 (\hbar c)^3}{g_s} A_{00}^2 \, n \, .
\end{equation}

\subsubsection{Energy density choice}

We can also choose $T_\mathrm{ref}$ as the temperature of the blackbody which has the same energy density (``energy density choice"). From~\eqref{rhofromc01} applied with a blackbody ($c_0 = 1/A_{00}$, $c_1 = 0$), we see that this corresponds to
\begin{equation}
\label{eq:Tref_equalenergy}
(\kB T_{\rm ref})^4 = \frac{2 \pi^2 (\hbar c)^3}{g_s} \frac{A_{11} A_{00}^2}{(-A_{10})} \, \rho \,.
\end{equation}
Note that $(-A_{10})/(A_{11} A_{00}^2)$ is $\pi^4/15$ for bosons and $(7/8) (\pi^4/15)$ for fermions. Once this temperature is chosen, from~\eqref{rhofromc01} we see that the coefficients describing the spectrum must satisfy
\begin{equation}
c_0  + \frac{A_{00}}{(-A_{10})}c_1 = \frac{1}{A_{00}}\,.
\end{equation}
However we are not free to choose $c_0$ since it is constrained by $n$ given the choice of reference temperature, thanks to~\eqref{nfromc0}.

\subsubsection{Graybody choice}

Another simple choice of reference consists in imposing $c_1=0$. We see from Eq.~\eqref{eq:nandrhofromci} that in that case the ratio $\rho/n$, which is the average energy per particle, does not depend on $c_0$. In other words, this choice consists in picking a reference temperature such that the associated blackbody has the same average energy per particle. This gives the condition 
\begin{equation}
\kB T_{\rm ref} = \frac{A_{11}}{(-A_{10})}\frac{\rho}{n}\,,
\label{Tref_equal_average_energy}
\end{equation}
and with~\eqref{nfromc0} this also fixes $c_0$ to
\begin{equation}
c_0 = \frac{2 \pi^2 (\hbar c)^3}{g_s} \frac{A_{00} (-A_{10})^3}{A_{11}^3}\frac{n^4}{\rho^3}\,.
\end{equation}
The expansion is then of the type
\begin{equation}
f(E) = \frac{1}{{\rm e}^{\bref E} \pm 1}\left[c_0 A_{00} + \sum_{i=2}^\infty c_i P_i(\bref E)\right]\,.
\end{equation}

Note that with this choice, the condition $c_1=0$ implies that the zeroth order truncation is the unique graybody (with $g=1-c_0 A_{00}$) which has the same number and energy densities as the full spectrum. In reference to that specific zeroth order truncation spectrum, we refer to this choice of reference temperature as the ``graybody choice".

\subsubsection{Graphical interpretation}

\begin{figure}[!ht]
    \centering
    \includegraphics[width=\columnwidth]{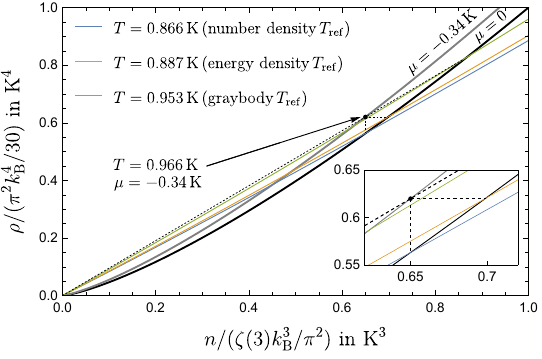}
    \caption{Illustration of the determination of the various $T_{\rm ref}$ for a given thermalized spectrum (solid point) of fermions. Solid lines correspond to lines of thermal spectra. When $\mu$ is specified, $T$ is varied along the line (black and gray lines), and conversely when $T$ is specified to the three $T_{\rm ref}$ defined in the text, $\mu$ is varied (colored lines). Dashed lines correspond to the directions of projection, with a vertical projection for constant number density, an horizontal projection for constant energy density, or a line of fixed ratio $\rho/n$ for constant average energy per particle.}
    \label{TrefChoice}
\end{figure}

It is convenient to illustrate these possible choices for $T_{\rm ref}$ in a number density/energy density plane, as in Fig.~\ref{TrefChoice}. Indeed, there is a one-to-one correspondence between a pair $(n,\rho)$ and a choice of $(T,\mu)$ in a thermal spectrum~\eqref{DefF}. Choosing a reference temperature amounts to choosing a method to project a point in the plane onto the curve of blackbodies (defined by $\mu=0$). The number density and energy density reference temperature are simply the two natural projections onto the blackbody curve, namely the vertical (constant $n$) and horizontal (constant $\rho$) projections. Furthermore, there is also a unique graybody spectrum [identified with $T,g$ as in~\eqref{DefGrayBody}] associated to a pair $(n,\rho)$. Hence the graybody temperature is a direct identification using both mappings in the plane. It also corresponds to the projection on the blackbody curve through the line which crosses the origin of the plane as all graybodies at a given $T$ but different $g$ have the same ratio $\rho/n$.

\begin{figure*}[!ht]
    \includegraphics[width=\columnwidth]{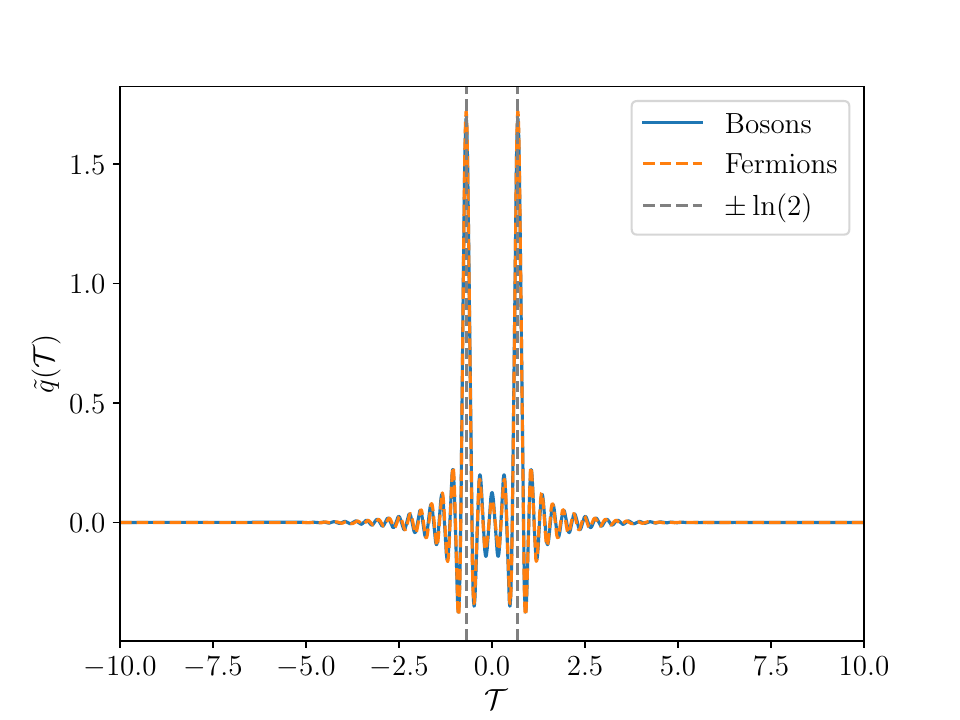}
    \includegraphics[width=\columnwidth]{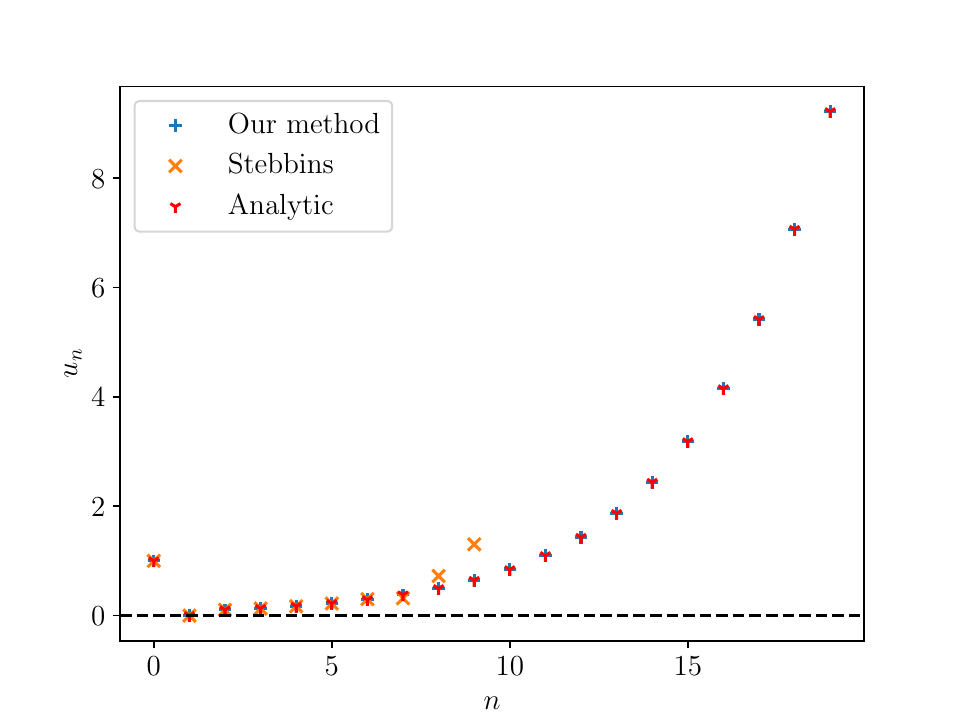}
    \caption{{\it Left:} Reconstruction of $\tilde{q}(\mathcal{T})$ for the spectrum \eqref{example_spectrum} of either bosons or fermions, using $\alpha = 10^{-25}$ in the regularization~\eqref{Wiener_regularization}. The Dirac functions of the analytical result~\eqref{example_spectrum_analytical_transform} are marked by the vertical dashed lines. 
    {\it Right:} Moments of the logarithmic temperature transform of the spectrum \eqref{example_spectrum}, calculated for the bosonic case. Our method and Stebbins' method are very close the analytic result \eqref{eq:centered_moments_twoT} at low $n$, but our implementation of Stebbins' method breaks down around $n \sim 7$ whereas our method works for larger $n$.
    }
    \label{fig:Tcomb_T_transform}
\end{figure*}

\section{Idealized distorted spectra}\label{SecIdeal}

Once a reference temperature $T_{\rm ref}$ is chosen, a spectrum can be expressed as a sum of a blackbody at that temperature and a distortion. That is for any spectrum $f(E)$ we define
\begin{equation}
\label{eq:distorted_spectrum}
E^2f(E) = E^2\bb(\bref E) + E^2 \delta f(E)\,.
\end{equation}
In this section, we consider idealized distortions and we will compare the reconstructions from a finite number of moments obtained as described in Sec.~\ref{SecLTT} for the LTT and Sec.~\ref{SecNLTT} for the NLTT, or with a finite number of coefficients in the OPE method described in Sec.~\ref{SecOPE}.

\subsection{Gray spectrum}

A gray spectrum at temperature $T$, given by~\eqref{DefGrayBody}, is directly and exactly described by $g \neq 0$ in the LTT method, with all $u_n=0$ ($n\geq2$), or equivalently by $\tilde{g}=g$ and all $\tilde{u}_n=0$ ($n \geq 2$). In the OPE, it is fully captured by the value of $c_0=(1-g)/A_{00}$ with all $c_i=0$ for $i \geq1$ if the reference temperature is the one of the gray spectrum.

\subsection{Incorrect reference temperature}

In the LTT description of distortions, all the central moments vanish for a pure blackbody at temperature $T$ and the spectrum is completely characterized by the value of $\overline{\cal T}=\widetilde{\overline{\cal T}}=\ln(T/T_0)$. However when using the OPE description, we must choose a reference temperature $T_{\rm ref}$, and if it differs from $T$ when performing the OPE, we might interpret the coefficients as the distortion since
\begin{equation}
E^2f(E) = E^2\bb(\bref E)+E^2\left[\bb(\beta E)-\bb(\bref E)\right]\,.
\end{equation}
Since when using the correct reference temperature the coefficients can only be $c_0=1/A_{00}$ and $c_i=0$ for $i \geq 1$,  the coefficients with the incorrect reference temperature can be deduced from~\eqref{MagicFlow1}, that is
\begin{equation}
\left.c_i\right|_{T_{\rm ref}} = \frac{1}{A_{00}}\sum_{p=0}^i A_{ip} (A^{-1})_{p0} \left(\frac{T}{T_{\rm ref}}\right)^{3+p}\,,
\end{equation}
which can be further simplified with~\eqref{PureMagic}.
One can also directly use \eqref{coeffs_calc}, noting that for this particular distribution we have $f_{p+2} = I_{p+2}/\beta^{3+p}$, and then use Eq.~\eqref{PureMagic}. If we restrict to terms linear in $\delta T = T - T_{\rm ref}$, this reduces to
\begin{align}
\left.A_{00} c_0\right|_{T_{\rm ref}} &\simeq 1 + 3 \, \delta T/T\,,\label{cidT1}\\
\left.c_i\right|_{T_{\rm ref}} &\simeq \sum_{p=0}^i A_{ip} I_{p+2} (3+p)\delta T/T\,,
\end{align}
where the value of $A_{00}$ is given in~\eqref{ValueA00}.

\subsection{Toy model with two blackbodies}

\begin{figure*}[!ht]
    \centering
    \includegraphics[width=\columnwidth]{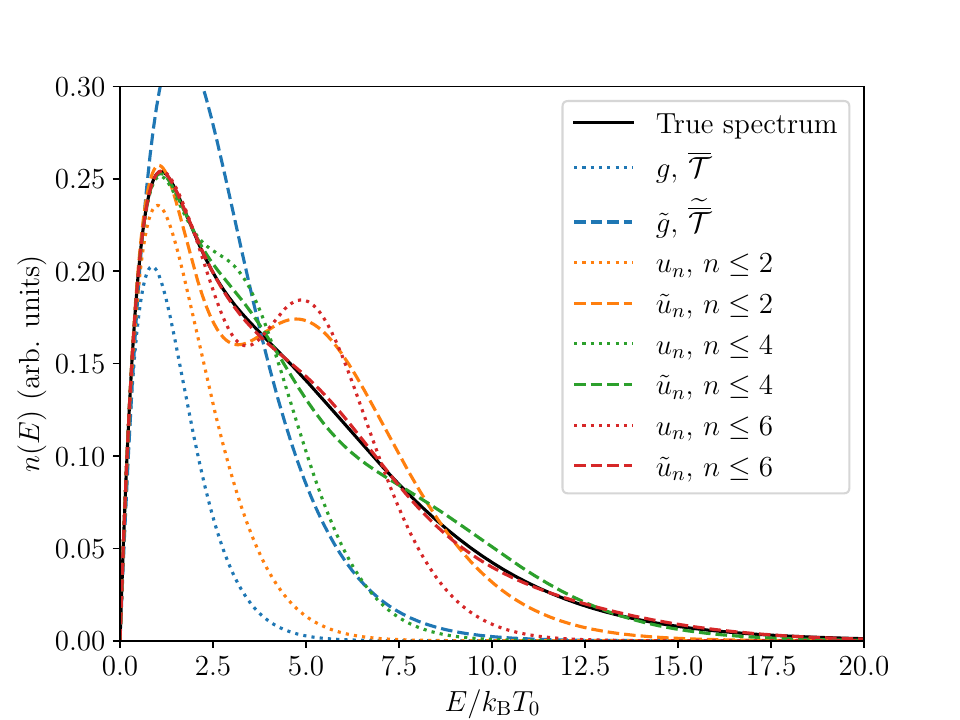}
    \includegraphics[width=\columnwidth]{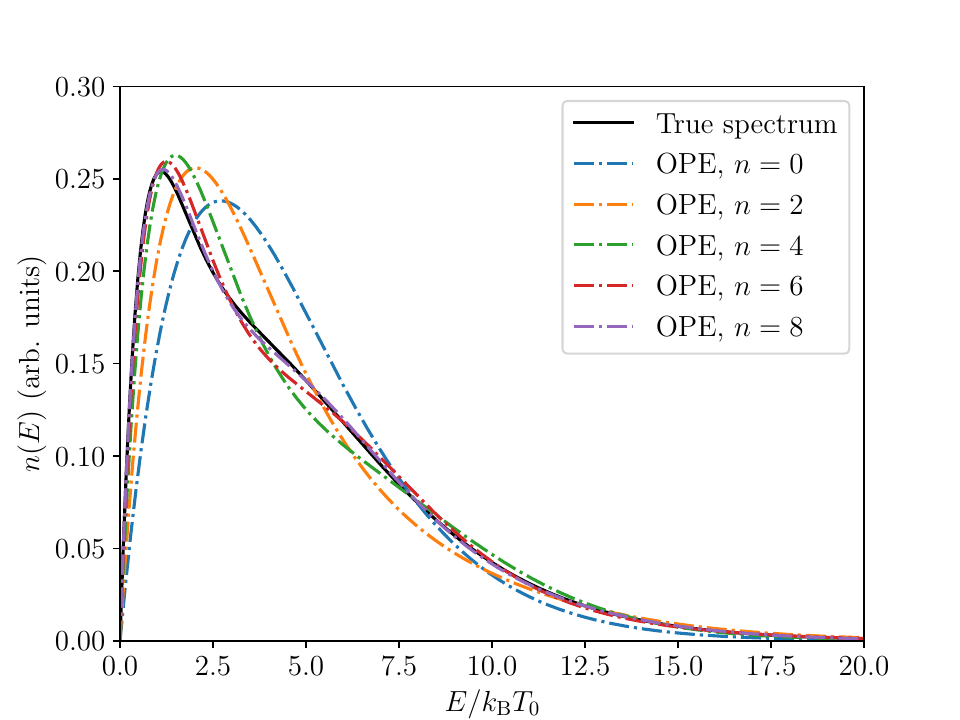}
    \caption{{\it Left: }Reconstructions of the number density distribution of the spectrum~\eqref{example_spectrum} in the bosonic case using the LTT and Eq.~\eqref{eq:fnu_from_moments} (dotted lines), or the NLTT and Eq.~\eqref{RecconstructionFromTilde} (dashed lines), truncated to various orders. The order 0 truncation is labeled as $g, \, \overline{\mathcal{T}}$ (resp. $\tilde{g}, \, \widetilde{\overline{\mathcal{T}}}$) to highlight that it is the graybody spectrum with the corresponding grayness and temperature.
    {\it Right: }OPE of the spectrum~\eqref{example_spectrum} in the bosonic case for various orders. The reference temperature corresponds to the graybody choice~\eqref{Tref_equal_average_energy}. We see that the approximations quickly converge toward the true spectrum.}
    \label{fig:rec_number_spectrum_from_moments}
\end{figure*}

Let us consider a toy model spectrum made of a superposition of two blackbodies with different temperatures, defined by
\begin{equation}
    f(E) = \bb\left(\frac{E}{\kB T_0/2}\right) +\frac{1}{2^4}\bb\left(\frac{E}{2 \kB T_0}\right)\,,
    \label{example_spectrum}
\end{equation}
where $T_0$ is an arbitrary temperature, which we also take as the reference temperature for the temperature transform [see Eq.~\eqref{eq:logvariables}]. The advantage of this spectrum is that the LTT (and NLTT) have analytic forms which are
\begin{subequations}
\begin{align}
    \tilde{q}(\mathcal{T}) &= \frac{1}{2^2} \delta\left(\mathcal{T}-\ln \frac{1}{2}\right) + \frac{1}{2^2}\delta\left(\mathcal{T} - \ln 2\right)\,,
    \label{example_spectrum_analytical_transform}\\
    q(\mathcal{T}) &= \delta\left(\mathcal{T}-\ln\frac{1}{2}\right) + \frac{1}{2^4}\delta\left(\mathcal{T} - \ln 2\right)\,,
\end{align}
\end{subequations}
hence it can be used to test our procedure to numerically reconstruct the logarithmic temperature transform.

Using the regularization parameter $\alpha = 10^{-25}$ in~\eqref{Wiener_regularization}, we calculate the NLTT $\tilde{q}(\mathcal{T})$. The left panel of Fig.~\ref{fig:Tcomb_T_transform} shows our result, both for bosons and fermions. The reconstruction recovers the expected analytical result reasonably well, with the sharpness of the peaks controlled by the smallness of $\alpha$. We also check that we recover the original spectrum using Eq.~\eqref{logT_convolution_tilde}.

The moments of the temperature transform are also known analytically and are
\begin{equation}
    q_n = (\ln 2)^n \left[ (-1)^n + \frac{1}{2^4} \right],\,\,  \tilde{q}_n = \frac{(\ln 2)^n}{2} \frac{\left[1+(-1)^n\right]}{2}\,.
\end{equation}
Using \eqref{Relqu}, we get the central moments
\begin{equation}
\label{eq:centered_moments_twoT}
    u_n = \frac{16}{17} \left(\frac{2 \ln 2}{17}\right)^n \left[(-1)^n + 16^{n-1}\right] \, .
\end{equation}
In our method, we compute the $u_n$ in the following way. From $\tilde{q}(\mathcal{T})$, we calculate the moment-generating function~\eqref{moment_generating_function_imag} to obtain the moments $\tilde{q}_n$ via~\eqref{moments_from_genfunction}. We deduce the LTT moments $q_n$ through~\eqref{moment_relation_series}, and finally obtain $u_n$ from Eq.~\eqref{Relqu}. On the right panel of Fig.~\ref{fig:Tcomb_T_transform}, we plot the central moments obtained with this procedure and compare them with those obtained with Stebbins' procedure~\cite{Stebbins:2007ve} and the analytic prediction~\eqref{eq:centered_moments_twoT}. Both are very close to the expected analytic results at low $n$. However, for larger $n$ values our Fourier method still works well whereas implementing Stebbins' procedure becomes extremely challenging. Although this procedure is in principle exact, we find that in practice it runs into numerical difficulties that heavily amplify numerical errors for large values of $n$. 

The initial spectrum is in principle reconstructed from its moments with~\eqref{eq:fnu_from_moments}, but this can only be approximate when we truncate the infinite sum on moments, which we illustrate in Fig.~\ref{fig:rec_number_spectrum_from_moments}. When comparing the reconstruction of the number density $n(E)$, the NLTT performs better than the LTT as seen on the left panel of Fig.~\ref{fig:rec_number_spectrum_from_moments}. For comparison, we also decompose the spectrum using the OPE. When choosing the graybody reference temperature, the decomposition converges rapidly to the exact spectrum, as illustrated in the right panel of Fig.~\ref{fig:rec_number_spectrum_from_moments}. We find that the OPE does not work well if we instead try to use the number density choice, which shows the importance of choosing a suitable $T_\mathrm{ref}$ for each spectrum. We recall that the choice of reference temperature $T_0$ in the log-temperature-transform approaches~\eqref{eq:logvariables} is much less important than the choice of $T_\mathrm{ref}$ for the OPE, since the central moments used for the reconstruction are independent of $T_0$ [see discussion after Eq.~\eqref{eq:fnu_from_moments}].

\subsection{$y$-type distortion}
\label{subsec:ydistortion}

\begin{figure*}[!ht]
    \centering
    \includegraphics[width=\columnwidth]{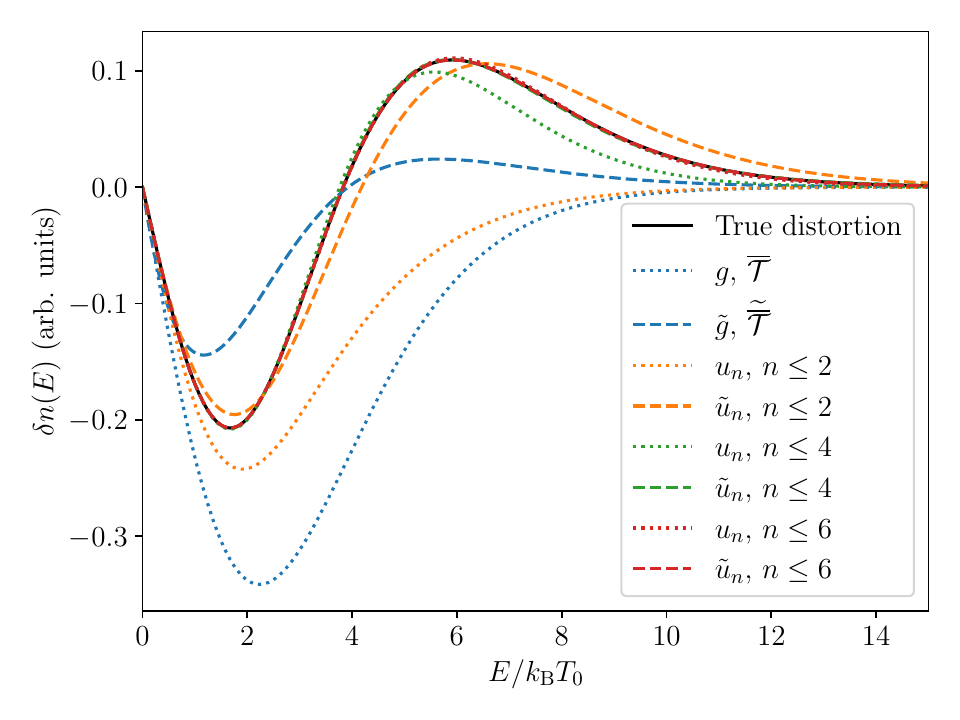}
    \includegraphics[width=\columnwidth]{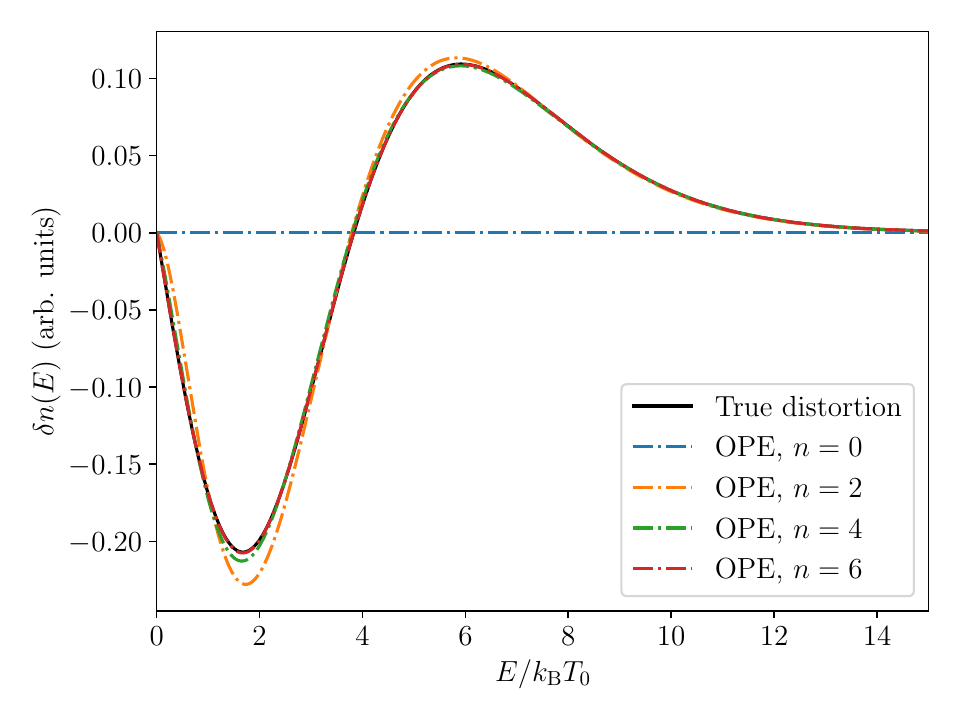}
    \caption{{\it Left: }Reconstructions of the number density distribution of a $y$ distortion spectrum (i.e. the true spectrum minus the undistorted blackbody spectrum) of bosons with $y=0.1$, using the LTT and Eq.~\eqref{eq:fnu_from_moments} (dotted lines), or the NLTT and Eq.~\eqref{RecconstructionFromTilde} (dashed lines), truncated to various orders. The order 0 truncation is labeled as $g, \, \overline{\mathcal{T}}$ (and $\tilde{g}, \, \widetilde{\overline{\mathcal{T}}}$) to highlight that it is the graybody spectrum with the associated grayness and temperature. We see that both methods work well, but the NLTT method converges faster.
    {\it Right: }OPE of the same bosonic $y$ distortion for various truncation orders. The reference temperature corresponds to the number density choice~\eqref{Trefnrelation}, which is especially natural for $y$ distortions because they do not alter the number density. We see that the approximations quickly converge toward the true spectrum.}
    \label{fig:rec_y_number_spectrum}
\end{figure*}

The general shape of $y$-type distortions stems from the Kompaneets equation~\cite{Kompaneets1957,Pitrou:2019hqg} evaluated in the single scattering approximation. It is given by
\begin{equation}\label{DefSZ}
f_y(E) = \bb(\beta E) + y \left(\frac{\partial^2}{\partial \ln E^2}+3\frac{\partial}{\partial \ln E}\right)\bb(\beta E) \,.
\end{equation}
By construction, if the reference temperature is the temperature $T=1/\kB \beta$ of the previous spectrum, the coefficients $c_i$ are proportional to $y$ except $c_0=1/A_{00}$ which does not depend on $y$ since this distortion preserves the number density. Evaluating~\eqref{coeffs_calc} via integration by parts and combining with~\eqref{PureMagic}, we obtain for $i \geq 1$ 
\begin{equation}
 c_i =y \times \sum_{q=0}^i A_{iq} \, q\,(q+3)\, I_{q+2}\,.
\end{equation}
Numerically, we find for bosons
\begin{align}
c_1 &\simeq 9.58458 y\,, & c_2 &\simeq 7.16155 y\,,\nonumber\\
c_3&\simeq 0.447752 y\,, & c_4 &\simeq -0.293001 y\,, \\
\intertext{and for fermions}
c_1 &\simeq  9.75912 y\,, & c_2 &\simeq 6.93075 y\,,\nonumber\\
c_3 &\simeq -0.248779  y\,, & c_4 &\simeq 0.115988 y\,.
\end{align}
The smallness of $\partial c_3/\partial y$ and $\partial c_4/\partial y$ ensures that restricting to $c_n$ with $n \leq 2$ provides a very good approximation of the $y$-distortion.

On the other hand, the expected LTT moments can be read by direct comparison of~\eqref{DefSZ} with~\eqref{eq:fnu_from_moments2} and we find $q_0=1$, $q_1=-3 y$, and $q_2=2y$ which leads to $g=0$ and
\begin{equation}
\begin{aligned}
\widetilde{\overline{\cal T}} &=\ln (T/T_0) -3y\,, \\
u_n &= (n-1)y(n-9y)(3y)^{n-2}\,.
\end{aligned}
\end{equation}
We checked that the method based on the NLTT (section~\ref{SecNLTT}) recovers these values with exquisite precision. Figure~\ref{fig:rec_y_number_spectrum} shows the reconstruction of a bosonic $y$ distortion with $y=0.1$ using the LTT, NLTT and OPE methods. For the LTT and NLTT methods, the reference temperature $T_0$ is the one of the undistorted spectrum, and since the $y$ distortion conserves the number density, it corresponds to a ``number density choice" $T_\mathrm{ref}$~\eqref{Trefnrelation}. This same $T_\mathrm{ref}$ is used for the OPE method, such that $c_0 = 1/A_{00}$ and the $n=0$ truncation of the distorted part of the spectrum is exactly zero (see dash-dotted blue line on the right panel). We see that, in this case, all three methods work very well, although the LTT method converges more slowly than the others.

\subsection{$\mu$-type distortion}
\label{subsec:mudistortion}

A thermal spectrum with a chemical potential is not strictly speaking a distorted spectrum. Nonetheless, we can decompose its spectral shape in the form~\eqref{eq:distorted_spectrum}. A $\mu$-type spectrum is of the form
\begin{equation}
f_\mu(E) = \bb\left(\frac{E-\mu}{\kB T}\right)\,.
\end{equation}
It is customary to characterize it with its reduced chemical potential $\xi \equiv \mu/\kB T$, and natural to choose $T_{\rm ref}=T$. 

From the definition of integrals~\eqref{DefJn}, the energy moments, and thus the $c_i$ when using~\eqref{coeffs_calc}, have the simple analytic form
\begin{equation}\label{cipolylog}
\begin{aligned}
\beta^{p+1}_{\rm ref} f_{\mu,p} &= p! J_p(\xi) \, , \\ c_i &= \sum_{q=0}^i A_{iq} (q+2)!J_{q+2}(\xi)\,.
\end{aligned}
\end{equation}

We must now distinguish the case of fermions and bosons.
\paragraph{Fermions} Using $\partial_\xi{\rm Li}_s(\pm{\rm e}^\xi)={\rm Li}_{s-1}(\pm{\rm e}^\xi)$, and $-{\rm Li}_s(-1) = \eta(s)$ and the definition~\eqref{DefJn}, we recover the series expansion of $J_q(\xi)$ for fermions which is
\begin{equation}\label{ExpansionJ}
J_q(\xi) = \sum_{k=0}^\infty \frac{\eta(q+1-k)}{k!} \xi^k\,.
\end{equation}
From~\eqref{cipolylog} we then deduce the expansion of the coefficients
\begin{equation}\label{cimagicxi}
c_i= \sum_{n=0}^\infty \left(\sum_{q=0}^i A_{iq} \frac{(q+2)!}{n!}\eta(q+3-n) \right)\xi^n \,.
\end{equation}
We obtain a very good approximation for $|\xi|<0.2$ for fermions when including terms of order $\xi^3$ in the previous expansion. The numerical values of the first coefficients in~\eqref{cimagicxi} are (with $1/A_{00}\simeq 1.34279$)
\begin{equation}
\begin{aligned}
c_0 &\simeq 1/A_{00} + 1.22501 \xi + 0.51620 \xi^2 + 0.12412 \xi^3\,,\\
c_1&\simeq 0.0968034 \xi + 0.121527  \xi^2 + 0.0720994 \xi^3\,,\\
c_2&\simeq -0.0628209  \xi -0.073359  \xi^2 -0.038081\xi^3\,,\\
c_3&\simeq 0.0355277\xi + 0.0368396  \xi^2 + 0.014439\xi^3\,,\\
c_4&\simeq -0.0177428 \xi -0.0145829 \xi^2 -0.00161364 \xi^3\,.
\end{aligned}
\end{equation}

\paragraph{Bosons} The distortion is only defined if $\xi\leq0$. The analytic expression~\eqref{cipolylog} is still correct but counterpart of the expansion~\eqref{ExpansionJ} is more involved (since $J_q(\xi)$ is not analytic anymore at $\xi=0$) and reads\footnote{See e.g.~(9.5) in Ref.~\cite{Wood92} or~(9.554) in Ref.~\cite{gradshteyn2007} using the relation between the polylogarithm and Lerch transcendent functions ${\rm Li}_s(z) = z \Phi(z,s,1)$.}
\begin{equation}
J_q(\xi) = \sum_{\substack{k=0 \\ k\neq q}}^\infty \frac{\zeta(q+1-k)}{k!} \xi^k + \frac{\xi^q}{q!}[H_q-\ln(-\xi)]
\end{equation}
where $H_q$ is the $q$-th harmonic number. When replaced~\eqref{cipolylog} we obtain a good numerical approximation for $|\xi|<0.2$ with (using $1/A_{00}\simeq 1.55052 $)
\begin{equation}
\begin{aligned}
c_0 &\simeq 1/A_{00} + 2.12178 \, \xi +0.644945 \, \xi^2 [3/2-\ln(-\xi)]\\
&\quad-0.107491 \, \xi^3  \,,\\
c_1 &\simeq -0.617741 \, \xi \\
&\quad + \xi^2 [1.82085 -  0.996685 (3/2 - \ln(-\xi))] \\
    &\quad + \xi^3 [0.166114 + 0.368982 (11/6 - \ln(-\xi))]\,,\\
c_2 &\simeq 0.577711 \, \xi \\
&\quad + \xi^2 [-2.92885 +     1.28254 (3/2 -  \ln(-\xi))] \\
    &\quad + \xi^3 [0.645299 -  0.975145 (11/6 -  \ln(-\xi))]\,,\\
c_3 &\simeq -0.522797 \xi + \xi^2 [4.01956 -     1.53019 (3/2- \ln(-\xi))] \\
    &\quad + \xi^3 [-2.10719 + 1.78013 (11/6 -  \ln(-\xi))]\,,\\
c_4 &\simeq 0.472838 \xi + \xi^2 [-5.07069 +     1.7519 (3/2 -  \ln(-\xi))] \\
    &\quad + \xi^3 [4.22125 -     2.76026 (11/6 -  \ln(-\xi))]\,.
\end{aligned}
\end{equation}

In both cases $c_0 = 1/A_{00}$ for the undistorted spectrum as expected, and $\partial c_0/\partial \xi = 2 I_1/\sqrt{I_2}$. Knowing this, it is possible to combine a $\mu$-distortion with a temperature distortion such that at linear order the number density of the combined distortion vanishes~\cite{Chluba:2013pya}. Using~\eqref{cidT1} with~\eqref{ValueA00}, we need to choose $\delta T/T = -(2/3)(I_1/I_2) \xi$. This is $\delta T/T =-(2/9)\zeta(2)/\zeta(3)\xi\simeq -0.304096 \xi$ for fermions and $\delta T/T =-(1/3)\zeta(2)/\zeta(3)\xi \simeq -0.456144 \xi$ for bosons.

\begin{figure}[!ht]
    \centering
    \includegraphics[width=\columnwidth]{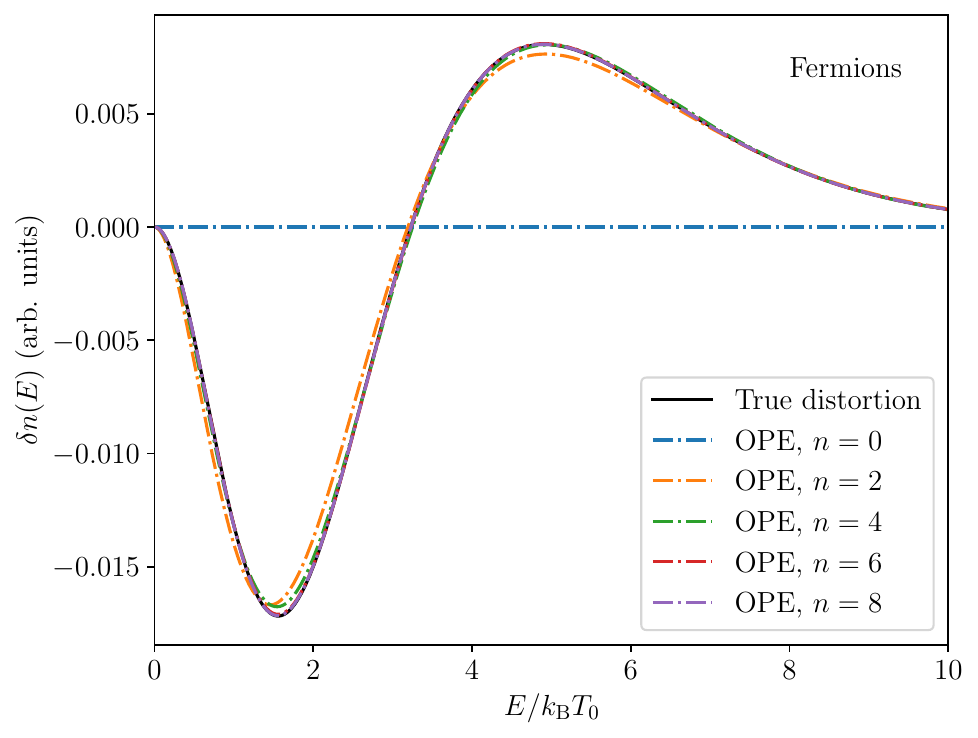}
    \includegraphics[width=\columnwidth]{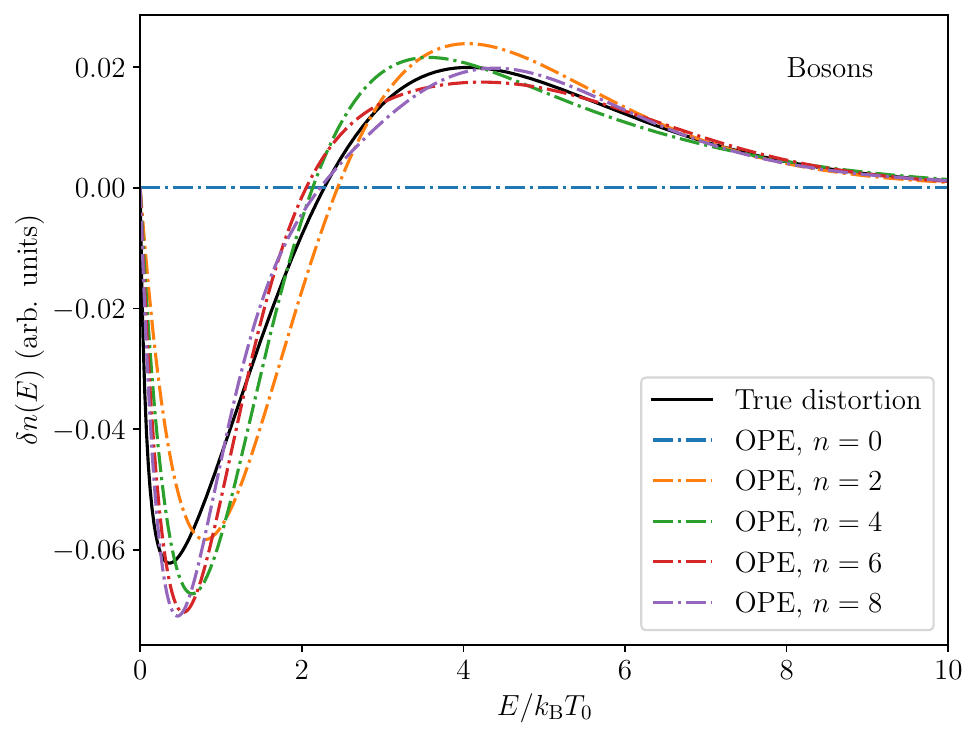}
    \caption{{\it Top:} OPE of the distortion of a fermionic spectrum with a chemical potential $\xi = \mu/\kB T = -0.1$ with respect to a thermal spectrum with the same number density, but a different temperature, in analogy to the ``$M(\nu)$" $\mu$-distortions from Ref.~\cite{Chluba:2013vsa}. The reference temperature corresponds to the number density choice. We see that the method works very well in the fermionic case.
    {\it Bottom:} same, but for a bosonic distortion with the same $\xi$. We see that the method converges much more slowly than in the fermionic case. \vspace{-0.5cm}}
    \label{fig:muF_dist_polydecomp}
\end{figure}

We show in Fig.~\ref{fig:muF_dist_polydecomp} that a fermionic spectrum with a chemical potential can be well described by the OPE method. However, we also see in the figure that the OPE method does not converge very well for $\mu$-distortions of bosons. This is still an improvement over the temperature transform methods which all fail in that case. Indeed, it was shown in Ref.~\cite{Pitrou:2014ota} that the LTT method does not work for $\mu$-type distortions of bosons, and we recover numerically that this is indeed the case for both bosons and fermions, even when using the improved NLTT method.

As it is easier to implement numerically, and since it converges in more situations than the log-temperature transforms, we advocate for a use of the OPE in a wide range of situations where distorted fermionic or bosonic spectra are involved. To illustrate how it can be used in practice, we apply the OPE in several cosmological contexts in the following section.

\section{Distorted spectra in cosmology}
\label{SecApplications}

\begin{figure*}[!ht]
    \centering
    \includegraphics[width=0.99\textwidth]{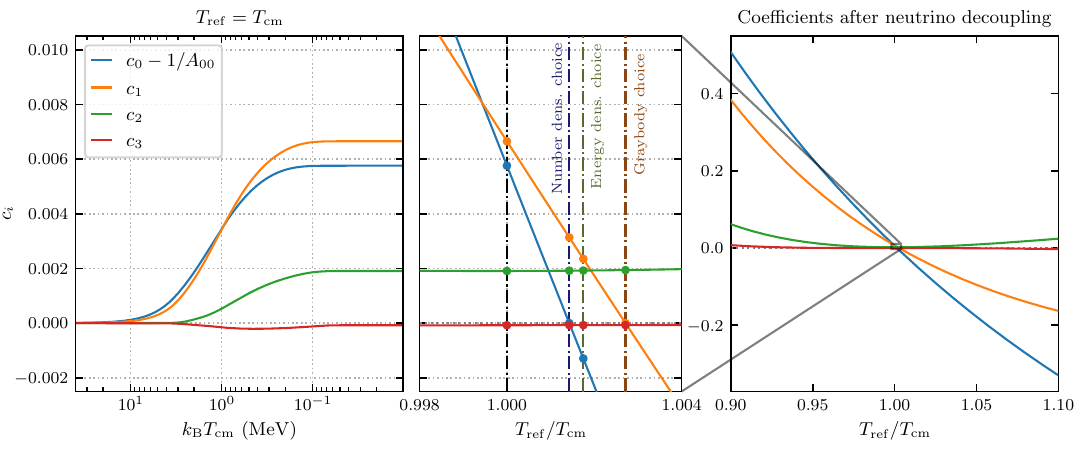}
    \caption{\emph{Left:} polynomial decomposition of the distribution function of $\nu_e$ throughout the neutrino decoupling epoch, for a reference temperature $T_\mathrm{ref}=T_\mathrm{cm}$. The neutrino spectrum is taken from Ref.~\cite{Froustey:2020mcq}. \emph{Right:} flow of the $c_i$ coefficients describing the frozen-out $\nu_e$ spectra (for $\kB T_\mathrm{cm} \ll 1 \, \mathrm{MeV}$) when changing the reference temperature. In the middle panel, we zoom in on the various $T_\mathrm{ref}$ choices discussed in Sec.~\ref{subsec_Tref}.}
    \label{fig:coefficients_CnuB}
\end{figure*}

We now proceed to apply the polynomial formalism introduced in Sec.~\ref{SecPE} to various situations where distorted spectra arise in cosmology and astrophysics. In general, the decoupling of any species, which is inherently an out-of-equilibrium process, leaves some nonthermal distortions on the spectra of the decoupled species. This happens for the cosmic neutrino background (C$\nu$B), when the temperature of the Universe is $\kB T \sim 1 \, \mathrm{MeV}$, and also for the cosmic microwave background (CMB), for $\kB T \sim 0.25 \, \mathrm{eV}$. More generally, nonstandard scenarios can lead to much larger distortions of these relic backgrounds. The measurements of cosmological observables can then be used to constrain such nonstandard scenarios. In this section, we show that the polynomial method we introduced can describe the small, expected, distortions in the standard decoupling scenario, but they also provide a convenient framework to constrain nonstandard scenarios.

\subsection{Primordial neutrino spectrum}
\label{subsec:CnuB}

\subsubsection{Standard neutrino decoupling scenario}

The standard cosmological model predicts the presence of a background of nearly-thermal neutrinos known as the cosmic neutrino background (C$\nu$B). Describing the decoupling of neutrinos from the electromagnetic plasma of photons, electrons and positrons for a temperature $\kB T \sim 1 \, \mathrm{MeV}$ requires solving the Boltzmann equations for the neutrino distribution functions throughout the decoupling era~\cite{Dolgov:1997mb}. Furthermore, if one wants a permille prediction of the standard value of the effective number of relativistic species $\Neff$, a number of phenomena must be taken into account: flavor oscillations, QED corrections to the plasma thermodynamics... Such precision calculations have been developed in the last decades, see, e.g.~\cite{Mangano:2005cc,Grohs:2015tfy,Akita:2020szl,Froustey:2020mcq,Bennett:2020zkv}.

The nonthermal distortions resulting from the out-of-equilibrium decoupling process are small (at the percent-level), such that our polynomial description should be extremely effective in describing the neutrino spectra. We note that an expansion of the spectral distortions on a set of orthonormal polynomials was previously used to compute neutrino decoupling in~\cite{Esposito:2000hi,Mangano:2001iu,Froustey:2019owm}, although the weight function did not include the factor $x^2$ compared to Eq.~\eqref{ScalarProduct_def}. Similarly to the difference between the LTT and the NLTT in Sec.~\ref{SecTT}, we choose to work at the level of the number density distributions. To illustrate the adequacy of our OPE approach, we use the electron neutrino distributions from Ref.~\cite{Froustey:2020mcq} and calculate the associated polynomial coefficients via Eq.~\eqref{coeffs_calc}. We show in the left panel of Fig.~\ref{fig:coefficients_CnuB} the evolution of the first three $c_i$ coefficients through the decoupling era. The x-axis shows the ``comoving temperature" $T_\mathrm{cm}$, which is the temperature neutrinos would have in the instantaneous decoupling approximation. It scales inversely with the scale factor $T_\mathrm{cm} \propto a^{-1}$. The reference temperature is also this comoving temperature, $T_\mathrm{ref}=T_\mathrm{cm}$. Because of this choice, after neutrino decoupling, when the only change in the spectrum is due to redshifting, the coefficients do not change anymore, consistent with the discussion after Eq.~\eqref{MagicRedshift}. On the right panel, we show how these coefficients describing the frozen-out spectrum ($\kB T_\mathrm{cm} \ll 1 \, \mathrm{MeV}$) vary when using different reference temperatures. To account for the fact that all physical temperatures decrease with the expansion of the Universe, we consider the ratio of $T_\mathrm{ref}$ with $T_\mathrm{cm}$. Since the distortions of the C$\nu$B are at the percent-level, the three choices discussed in Sec.~\ref{subsec_Tref} span a small range of values, on which we focus in the middle panel. We note that the ``energy density choice"~\eqref{eq:Tref_equalenergy} allows to define an unperturbed spectrum with the correct energy density, and thus corresponding to the correct $\Neff$. This choice was notably used in~\cite{Froustey:2019owm,Froustey:2020mcq} to distinguish various effects of neutrino distributions on BBN, but the remaining nonthermal distortions were not fitted into our OPE framework. We clearly see in the middle panel that the curve $c_0(T_\mathrm{ref}) - 1/A_{00}$ [resp. $c_1(T_\mathrm{ref})$] crosses 0 for the number density choice~\eqref{Trefnrelation} [resp. the graybody choice~\eqref{Tref_equal_average_energy}].

\subsubsection{Constraints on non-standard C$\nu$B distortions}

Few observational constraints exist on the spectrum of the C$\nu$B, potentially allowing for large distortions in non-standard cosmologies \cite{Cuoco:2005qr, Alvey:2021sji}. Besides, standard cosmology predicts that neutrino masses should have an effect on large-scale structure, but this effect has not been seen yet. In fact, recent results by DESI~\cite{DESI:2025zgx,DESI:2024mwx} show a $\sim 3 \sigma$ tension between the cosmological measurements and the minimum mass compatible with neutrino flavor oscillations~\cite{Elbers:2025vlz,2024JHEP...09..097C_negativenumass,Jiang:2024viw,Escudero:2024uea}, which motivates us to consider the possibility of non-standard features in the cosmic neutrino spectrum. Certain types of non-standard C$\nu$B distortions have been studied and constrained, such as neutrino $\mu$ distortions caused by a neutrino asymmetry \cite{Simha:2008mt,Shimon:2010ug,Castorina:2012md,Barenboim:2016shh,Oldengott:2017tzj,Escudero:2022okz,Froustey:2024mgf,Lattanzi:2024hnq,Domcke:2025lzg,Domcke:2025jiy}, neutrino $y$ distortions \cite{Barenboim:2024wek} and gray distortions \cite{Barenboim:2025vrc}. Here, instead of assuming a specific model, we want to characterize and constrain C$\nu$B distortions in a generic, model-independent way. The orthonormal polynomial expansion introduced in Sec.~\ref{SecPE} offers a natural framework to do so (see for instance~\cite{Bond:2024ivb} for another approach).

Here, we assume that all neutrinos and antineutrinos have the same spectrum, of the form~\eqref{Spectrum_decomposition}, but truncated at order 2 (i.e., $c_i = 0$ for $i \geq 3$), which is sufficient given the low constraining power that existing observations have on the C$\nu$B spectrum. We take the comoving temperature as the reference $T_\mathrm{ref} = T_\mathrm{cm}$, and assume constant coefficients as the temperature changes. For the standard C$\nu$B distortions, we have verified that setting the $c_i$ coefficients to their final asymptotic values (see left panel of Fig.~\ref{fig:coefficients_CnuB}) leads to a subdominant change in the primordial abundances. In addition, we must be careful and make sure that our distorted spectrum is physical. Indeed, a truncated OPE is not guaranteed to correspond to a physical spectrum.  We thus impose two additional conditions on the set of $\{c_i\}$: the distribution must be positive, and since neutrinos are fermions, it must be smaller than 1 to satisfy Pauli's exclusion principle. However, there are some cases where the spectrum only becomes negative at very high energies, where the spectrum is so heavily exponentially suppressed that this is irrelevant in practice. To allow this case, we actually only impose the condition $0 \leq f(E)\leq 1$ for $0 \leq E/\kB T_\mathrm{ref} \leq 20$, considering that unphysical values beyond this range are harmless features of the truncation.

\begin{figure*}[!ht]
    \centering
    \includegraphics[width=0.95\textwidth]{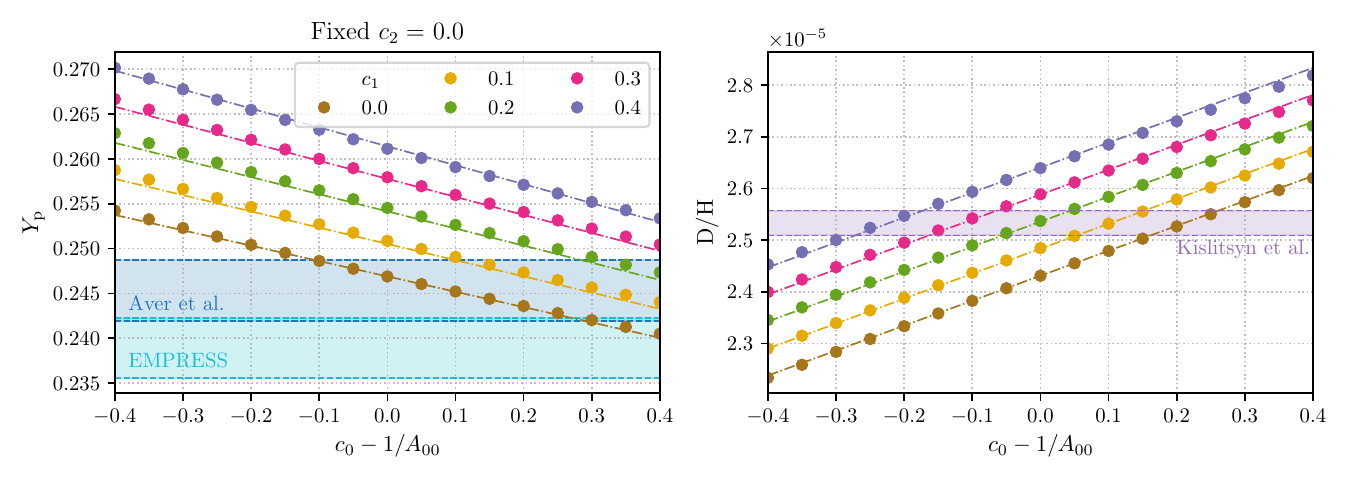}
    \includegraphics[width=0.95\textwidth]{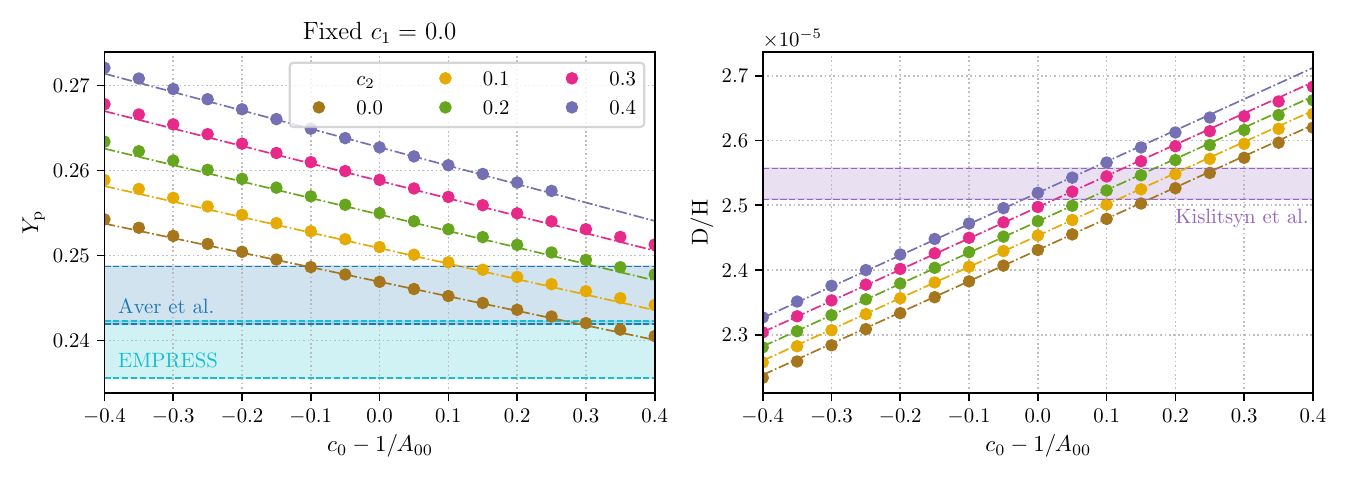}
    \caption{Primordial abundances of helium-4 (\emph{left}) and deuterium (\emph{right}) obtained with \primat, varying the OPE coefficients describing the neutrino spectra during the BBN epoch. The baryon abundance is fixed here at $\omega_b = 0.0224$. The top panels (resp.~bottom panels) show the dependence of $\Yp$ and $\DH$ at fixed $c_2 = 0$ (resp.~at fixed $c_1 = 0$). The dash-dotted lines correspond to the fitting functions~\eqref{eq:fit_Yp}--\eqref{eq:fit_DH}. The colored bands correspond to the measured spectroscopic values~\eqref{eq:Yp_mes}--\eqref{eq:Yp_empress}.}
    \label{fig:output_PRIMAT}
\end{figure*}

\paragraph{Experimental data}
The main observational constraints on the neutrino spectrum come first from the $N_\mathrm{eff}$ parameter, which measures (anti)neutrino total energy density relative to the photon energy density. Furthermore, since neutrino weak interactions affect the neutron-to-proton freeze-out ratio, BBN observations are also sensitive to the shape of the spectrum of $\nu_e$ and $\bar{\nu}_e$.

To constrain the coefficients, we employ BBN observations of deuterium from Kislitsyn \emph{et al.}~\cite{Kislitsyn:2024jvk} and of helium-4 from Aver \emph{et al.}~\cite{Aver2020}:
\begin{align}
    \Yp^\obs &= 0.2453 \pm 0.0034 &&\text{\protect\cite{Aver2020}} \, , \label{eq:Yp_mes} \\
    \DH^\obs &= (2.533 \pm 0.024) \times 10^{-5} &&\text{\protect\cite{Kislitsyn:2024jvk}} \, .\label{eq:DH_mes}
\end{align}
As a check, we also examine how our results change if we instead use helium-4 observations from EMPRESS~\cite{Yanagisawa:2025mgx}:
\begin{equation}
\label{eq:Yp_empress}
    \Yp^\obs\rvert_\mathrm{EMPRESS} = 0.2387^{+ 0.0036}_{-0.0031} \qquad \text{\protect\cite{Yanagisawa:2025mgx}} \, .
\end{equation}
In those expressions, $\Yp$ is the helium baryon fraction~\cite{Fields:2019pfx}. We also use the $N_\mathrm{eff}$ constraint from CMB~\cite{SPT-3G:2025bzu, ACT:2025tim, Planck:2018vyg} and CMB lensing~\cite{SPT-3G:2024atg,ACT:2023kun, Planck:2018lbu} that does not use information from BBN theory, as we are using non-standard BBN. Specifically, we use the constraint from the CMB-SPA combination with $\Yp$ as a free parameter\footnote{This is an approximation. In reality, we can model BBN in our non-standard scenario, so $\Yp$ is not a free parameter, and can be predicted from the coefficients describing the C$\nu$B distortions, as done in Ref.~\cite{Barenboim:2025vrc} for gray distortions. If one were to use such a procedure, the qualitative results would not change, but, quantitatively, the $\Neff$ constraint would become tighter, closer to the CMB-SPA $\Neff$ constraint with standard $\Yp$ in Ref.~\cite{SPT-3G:2025bzu}, $\Neff = 2.81 \pm 0.12$. This tight constraint also shows a $\sim 2\sigma$ preference for $\Neff$ being below its standard value. However, when combining with DESI, $\Neff$ is significantly pushed upward and this preference disappears. Due to the discrepancy between the two types of experiments, it is questionable whether it is justified to use such a tight constraint. Hence, we prefer to use the more conservative constraint that leaves $\Yp$ as a free parameter and is fully compatible with the standard model prediction.} in Table V in Ref.~\cite{SPT-3G:2025bzu}. As mentioned in that reference, one should not combine the constraints from the CMB-SPA combination with BAO data from DESI~\cite{DESI:2025zgx} because they are discrepant with each other. Hence, since we are dealing with an early-time description of the C$\nu$B, we use the CMB-SPA constraint:
\begin{equation}
    N_\mathrm{eff} = 2.99^{+0.22}_{-0.26} \qquad \text{\protect\cite{SPT-3G:2025bzu}} \, .
    \label{Neff_SPA_constraint}
\end{equation}
The baryon abundance $\omega_b=\Omega_b h^2$ is treated as a free parameter, but with a prior from the abundance inferred by the aforementioned CMB-SPA observations. More specifically, this is a Gaussian prior with $\omega_b = 0.022381 \pm 0.000093$ (see Table I in~\cite{SPT-3G:2025bzu}). In practice, we find that this prior constrains $\omega_b$ much more than BBN, so the $\omega_b$ posterior remains prior-dominated.

\paragraph{Predictions for cosmological observables}
To obtain the BBN abundances of helium-4 ($\Yp$) and deuterium ($\DH$) for each set of coefficients describing the C$\nu$B spectrum, we use a modified version of \texttt{PRIMAT}~\cite{Pitrou:2018cgg}. Similarly to what was done in~\cite{Barenboim:2025vrc}, we include the OPE-parametrized neutrino spectra, in order to take into account the changes in the neutrino energy density (with contributions from all three flavors) and in the weak $n \leftrightarrow p$ interconversion rates (modified by the $\nu_e, \bar{\nu}_e$ distributions). Finite nucleon mass and radiative corrections are computed at the level of the reference neutrino spectrum, and the spectral distortions enter at the Born approximation level (with a multiplicative factor describing radiative corrections~\cite{Pitrou:2018cgg,Froustey:2019owm}). Since we do not consider a specific scenario giving rise to the neutrino spectral distortions, we assume an instantaneous neutrino decoupling occurring sufficiently before BBN, such that the polynomial coefficients describe frozen-out, distorted neutrino spectra.

The output of \texttt{PRIMAT} for a subset of the parameter space is shown on Fig.~\ref{fig:output_PRIMAT}. The baryon abundance is $\omega_b = 0.0224$; the top (resp. bottom) panels show the results for $c_2 = 0$ (resp. $c_1 = 0$). On the bottom panels, we see that there are ``missing" data points for $c_2 = 0.4$ and $c_0 - 1/A_{00} \geq 0.3$: this is because the associated spectra do not satisfy our physicality criteria.

We have obtained the following fitting functions for the abundances of helium-4
\begin{align}
    \Yp &= \overline{\Yp} \, \mathrm{e}^{1.751 \Delta \omega_b} \nonumber\\
    &\quad - \left(0.01711 + 0.00993 \, c_1 + 0.01139 \, c_2\right) \left(c_0 - 1/A_{00}\right) \nonumber\\
    &\quad + \left(0.03623 - 0.01532 \, c_2\right) \, c_1 + 0.03955 \, c_2 \, ,  \label{eq:fit_Yp}
\end{align}
and of deuterium
\begin{align}
    10^5 \times \DH &= 10^5 \times \overline{\DH} \, \mathrm{e}^{-73.48 \Delta \omega_b} \nonumber\\
    &\quad + \left(0.4827 - 32.69 \, \Delta \omega_b\right) \left(c_0 - 1/A_{00}\right) \nonumber \\
    &\quad + \left(0.5230 - 34.75 \, \Delta \omega_b\right) \, c_1 \nonumber\\
    &\quad + \left(0.2205 - 14.28 \, \Delta \omega_b\right) \, c_2 \, , \label{eq:fit_DH}
\end{align}
with $\Delta \omega_b = \omega_b - 0.0224$, and $\overline{\Yp} = 0.2469$, $\overline{\DH} = 2.431 \times 10^{-5}$. These fits are shown with dash-dotted lines on Fig.~\ref{fig:output_PRIMAT}.

\begin{figure}[!ht]
    \centering    
    \includegraphics[width=0.96\columnwidth]{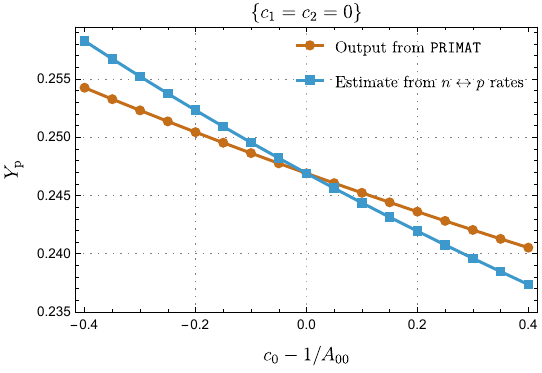} 
    \includegraphics[width=0.96\columnwidth]{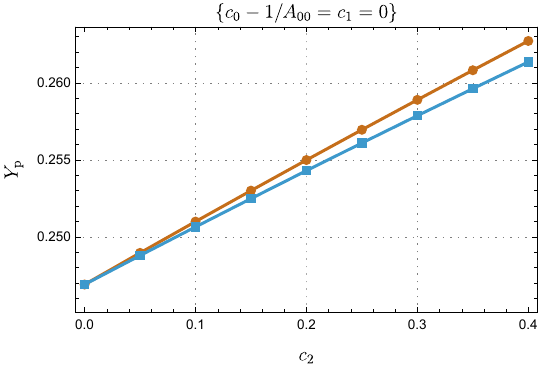}
    \caption{Helium-4 abundance obtained with \primat varying $c_0$ (top panel, setting $c_1 = c_2 = 0$) or $c_2$ (bottom panel, setting $c_0 = 1/A_{00}$ and $c_1 = 0$), in brown lines. The values correspond to what is depicted on Fig.~\ref{fig:output_PRIMAT}. We show in blue the estimate~\eqref{eq:Yp_proxy} based on neutron-to-proton freeze-out.}
    \label{fig:Yp_rates}
\end{figure}

We can understand the dependence of the primordial abundances on the coefficients of the C$\nu$B OPE, by scrutinizing the changes to the neutron/proton fraction at freeze-out. If we write $\Gamma_{n \to p}$ (resp.~$\Gamma_{p \to n}$) the rates of $n \to p$ (resp.~$n \to p$) reactions, as long as equilibrium is maintained the neutron fraction is given by
\begin{equation}
\label{eq:eqb_nfraction}
    \left.\frac{n_n}{n_n + n_p}\right|_\mathrm{eqb} = \frac{\Gamma_{p \to n}}{\Gamma_{p \to n} + \Gamma_{n \to p}} \ .
\end{equation}
Since most neutrons fuse into helium-4, a good proxy for $\Yp$ is given by twice
the neutron fraction at the onset of BBN. This value is modified by the $\{c_i\}$ with respect to the standard scenario in three ways. First, the freeze-out of $n \leftrightarrow p$ reactions occurs at different temperatures $T_\mathrm{FO}$, since the reaction rates and the Hubble expansion rate $H$ are modified by $\nu$ distributions. Second, the equilibrium value of $n_n/(n_n + n_p)$ is in itself modified through the changes of rates in Eq.~\eqref{eq:eqb_nfraction}. Third, the neutron fraction decreases between freeze-out and the onset of BBN (because of neutron decay, but see discussion in~\cite{Grohs:2016vef}). The extent of this decrease depends on the expansion rate, through the so-called ``clock effect"~\cite{Dodelson:1992km,Fields:1992zb,Froustey:2019owm}: a larger $\Neff$ changes the time-temperature relation to leave less time for neutron decay and this results in a larger neutron fraction at the onset of BBN, so a larger helium-4 abundance. In order to check that our results are physically sound, we show in blue lines on Fig.~\ref{fig:Yp_rates} a proxy for $\Yp$, namely
\begin{equation}
\label{eq:Yp_proxy}
    \left.\Yp\right\rvert_\mathrm{proxy} = \sigma \, \left.\frac{2 \Gamma_{p \to n}}{\Gamma_{p \to n} + \Gamma_{n \to p}}\right\rvert_{T_\mathrm{FO}} \, ,
\end{equation}
where the freeze-out temperature is estimated by numerically solving\footnote{There is no strict definition of the freeze-out temperature, which is usually expressed ``$\Gamma/H \sim 1$." But $\Gamma$ can be either the total reaction rate, or only the $n \to p$ one. Also, the exact numerical value on the right-hand side is not rigorously determined, see for instance the discussion in~\cite{Froustey:2019owm} where different choices allow to understand precisely the changes in freeze-out.} $\Gamma_{p \to n} + \Gamma_{n \to p} = 2 H(T_\mathrm{FO})$. The prefactor $\sigma \simeq 0.666$ allows us to crudely describe the decrease of the neutron fraction after freeze-out, it is set so that our estimate agrees with the actual value of $\Yp$ for $\{c_i\} = \{0\}$. 

On the top panel of Fig.~\ref{fig:Yp_rates}, we see that our estimate, although quite good, underestimates $\Yp$ as $c_0$ increases. This is consistent with the clock effect discussed above: since a larger $c_0$ corresponds to a larger $\Neff$, the prefactor $\sigma$ in~\eqref{eq:Yp_proxy} should be larger for large $c_0$, since less neutron decay takes place. On the bottom panel, we see that the discrepancy between the actual values and our estimate is smaller, which is expected since $\Neff$ does not depend on $c_2$. However, the limitations of our crude estimate of $T_\mathrm{FO}$ make a perfect agreement unattainable. Regarding deuterium, the same clock effect leads to a larger value of $\DH$ for larger $\Neff$, since most of BBN consists in a decrease of $\DH$ from its peak value~\cite{Froustey:2019owm}. This is consistent with the trends observed in Fig.~\ref{fig:output_PRIMAT}. The increase of $\DH$ with $c_2$, although smaller for instance than the increase with $c_1$, is also expected since with a larger $c_2$, the neutron fraction at the onset of BBN is larger, and a smaller proton fraction increases the value of $\DH$ at constant deuterium density. All in all, these results confirm our physical understanding of our BBN predictions.

\begin{figure*}[!ht]
    \centering
    \includegraphics[width=\columnwidth]{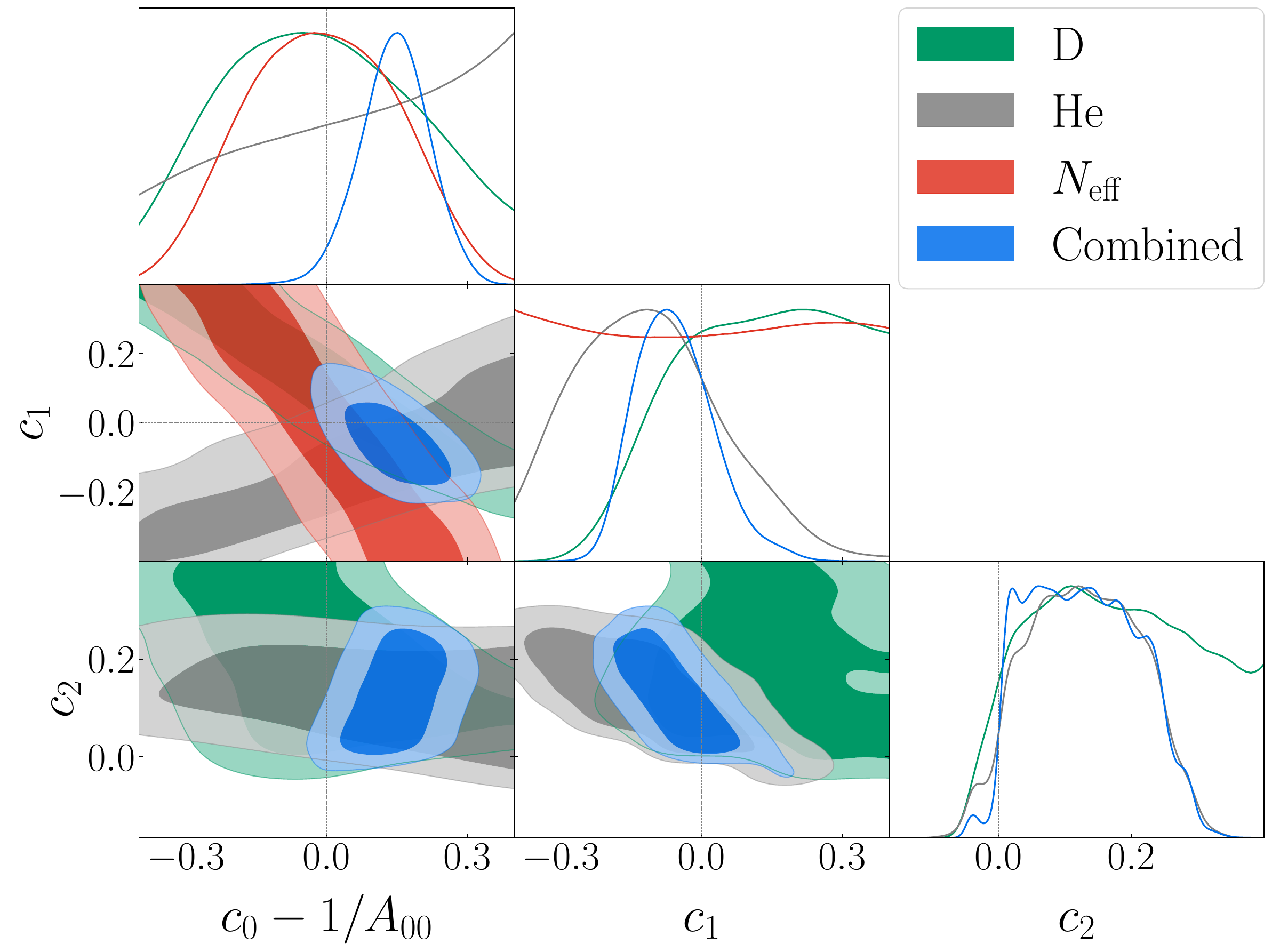}
    \includegraphics[width=\columnwidth]{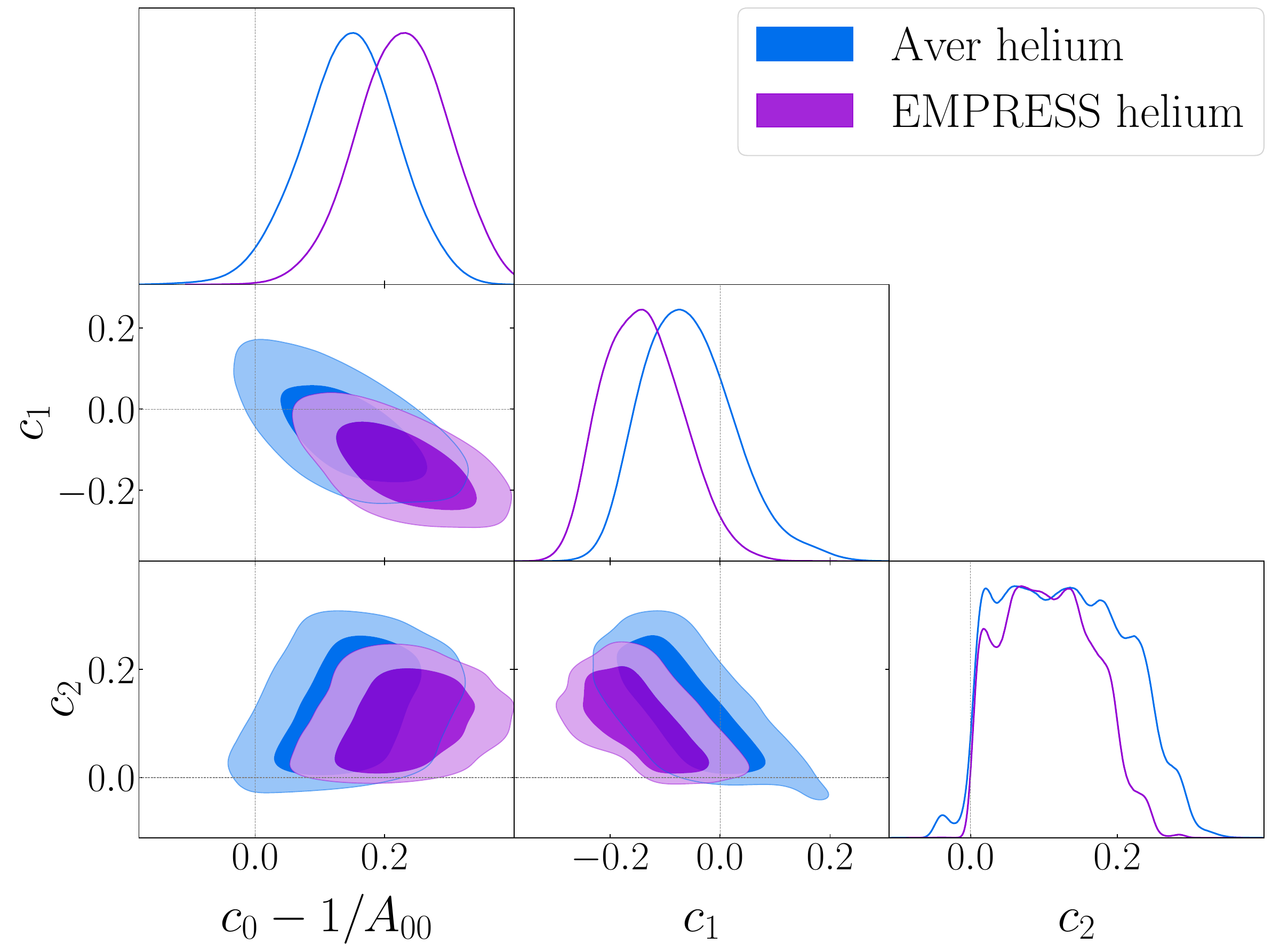}
    \caption{{\it Left:} Constraints on the first three coefficients of the OPE describing the C$\nu$B using observational data from BBN abundances of deuterium~\eqref{eq:DH_mes} and helium~\eqref{eq:Yp_mes}, as well as $N_\mathrm{eff}$ constraints from CMB, lensing and BAO~\eqref{Neff_SPA_constraint}. The figure shows each constraint separately, as well as the combined bound.
    {\it Right:} Constraints on the first three coefficients of the OPE describing the C$\nu$B using observational data from BBN and $N_\mathrm{eff}$ constraints. We compare the bounds obtained with helium observations from Aver \emph{et al.}~\cite{Aver2020} and EMPRESS~\cite{Yanagisawa:2025mgx}.
    }
    \label{fig:CnuB_constraints_compared}
\end{figure*}

To get $N_\mathrm{eff}$ for a given set of $c_i$, since $N_\mathrm{eff}$ only depends on the total energy density of the spectrum, we can employ~\eqref{rhofromc01} to deduce 
\begin{align}
    \frac{N_\mathrm{eff}}{N_\mathrm{eff,s}} &=  c_0 A_{00} + c_1  \frac{A_{00}^2}{(- A_{10})}\,, \\
\intertext{or equivalently}
    \frac{N_\mathrm{eff}}{N_\mathrm{eff,s}} &=  c_0 A_{00} + c_1 \left( A_{10} + A_{11} \frac{2700\zeta(5)}{7\pi^4} \right)
    \label{Neff_from_coefficients}\,,
\end{align}
where $N_\mathrm{eff,s}$ is the $N_\mathrm{eff}$ value corresponding to the reference spectrum with $c_0=1/A_{00}$, $c_{i>0}=0$, and $A_{00}$, $A_{10}$ and $A_{11}$ are the coefficients of $P_0$ and $P_1$, given in Appendix~\ref{ap:PolynomialCoefficients}. In our case,\footnote{This value corresponds to the standard model prediction if neutrino decoupling is assumed to be instantaneous, but QED corrections to the plasma thermodynamics are included \cite{Bennett:2019ewm}. Due to the instantaneous decoupling approximation, this differs by about 1~\% from the “true” standard value $N_\mathrm{eff}^\mathrm{std} = 3.044$~\cite{Akita:2020szl,Froustey:2020mcq,Bennett:2020zkv,Drewes:2024wbw}. This is not really an error, $N_\mathrm{eff,s}=3.010$ should instead be seen as a definition of the spectrum we consider ``undistorted", so that the standard spectrum has percent-level distortions with respect to this reference spectrum. In any case, this difference lies well below current experimental sensitivities~\cite{SPT-3G:2025bzu}.} $N_\mathrm{eff,s}=3.010$.

\paragraph{Constraints in the OPE framework}
Once we know how to obtain the observables for a given set of initial parameters, we derive constraints on the parameters from observational data using the MCMC code Cobaya \cite{Torrado:2020dgo}.

The left panel of Fig.~\ref{fig:CnuB_constraints_compared} shows the observational bounds on the coefficients. Note that the individual bounds (deuterium, helium and $\Neff$) are not stronger than the used prior, but the combined constraint is, as it can restrict the parameters to a bounded region within the prior volume that does not touch its boundaries. Since the coefficient $c_2$ does not affect the energy density of neutrinos, it is insensitive to the $\Neff$ constraint (see the red contours).

\begin{figure*}[!ht]
    \centering
    \includegraphics[width=\columnwidth]{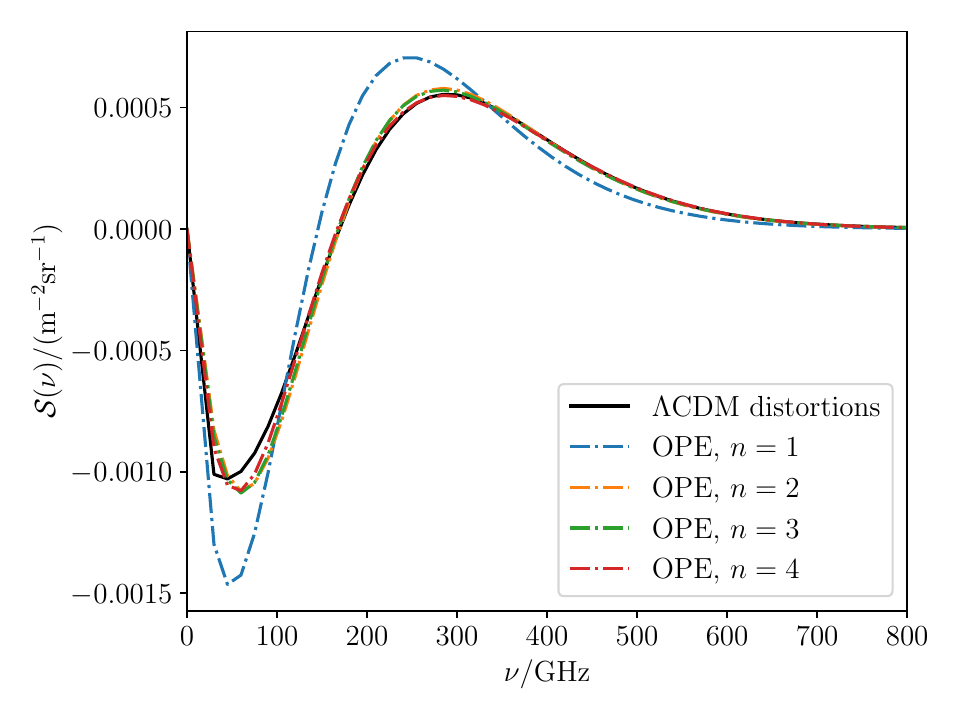}
    \includegraphics[width=\columnwidth]{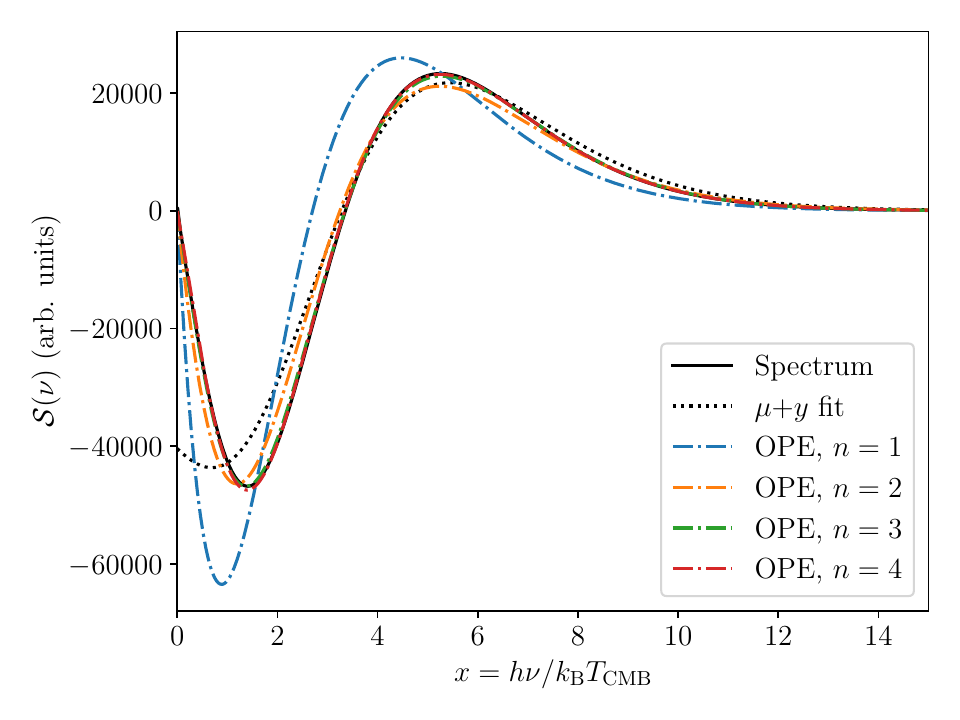}
    \caption{{\it Left: }Number radiance spectrum $\mathcal{S}(\nu)$ [related to the number density as $n_\mathrm{SD}(\nu) = (4\pi/c) \mathcal{S}(\nu)$] of CMB spectral distortions in the $\Lambda\rm{CDM}$ model and its polynomial decomposition, truncated to various degrees of the expansion.
    {\it Right: }Polynomial decomposition of the CMB distortion caused by a $\delta$-shaped energy injection at redshift $z=4\cdot 10^4$, when distortions are neither of the $\mu$ nor $y$ type. A fit of the distortion to a combination of $\mu$ and $y$ distortions is also shown. The polynomial decomposition can capture the details of the distortion more precisely, even with just a few coefficients. }
    \label{fig:standard_distortions_poly_decomp}
\end{figure*}

The right panel of Fig.~\ref{fig:CnuB_constraints_compared} shows how the combined bounds are modified depending on the chosen helium observations: Aver \emph{et al.} (blue) and EMPRESS (purple). Note that using the EMPRESS value slightly increases the preferred value of $c_0$, leading to a $\gtrsim2\sigma$ preference for $c_0-1/A_{00} > 0$. It also tightens the constraints on $c_1$ and $c_2$. This slight shift toward nonzero distortions is consistent with the preference for nonstandard scenarios to explain the EMPRESS value, such as a $y$-type distortion~\cite{Barenboim:2025vrc} or a nonzero electron (anti)neutrino asymmetry~\cite{Matsumoto:2022tlr,Burns:2022hkq,Escudero:2022okz,Froustey:2024mgf} (although we emphasize that in our analysis, the spectra of $\nu_e$ and $\bar{\nu}_e$ are identical and described by the same $\{c_i\}$).

We note that since our analysis uses the \primat determination of the deuterium abundance, there is slight tension in the nondistorted case with the measured abundance~\cite{Pitrou:2020etk}. This is clear in the right panels of Fig.~\ref{fig:output_PRIMAT}, where the points for $\{c_i\}=\{0\}$ are away from the measured value band. This tension, which is absent from the predictions of the BBN code \parthenope~\cite{Gariazzo:2021iiu}, can be traced to different choices made in \primat and \parthenope regarding the rates of the nuclear reactions $\mathrm{D}(d,n){}^3\mathrm{He}$ and $\mathrm{D}(d,p){}^3\mathrm{H}$~\cite{Pisanti:2020efz,Pitrou:2021vqr}. 

In principle, there are at least two other cosmological observables that can constrain the neutrino spectrum. One of them is the lithium abundance from BBN, whose measured value \cite{ParticleDataGroup:2024cfk, Sbordone_lithium_2010} is incompatible with the standard model prediction, which is known as the ``lithium problem"~\cite{Fields:2011zzb}. Even though a different neutrino spectrum could lead to a different lithium abundance, we have checked that this cannot solve the lithium problem without compromising the agreement with the other BBN abundances. Another source of constraint on the neutrino spectrum is the determination of the sum of neutrino masses. Existing upper bounds on this sum are starting to be in tension with lower bounds from oscillation experiments. When one measures the cosmological sum of neutrino masses, one is actually measuring the neutrino density at late times, when the neutrinos are non-relativistic. This late-time neutrino density depends on the number density of the neutrinos at early times, which, as we saw in Eq.~\eqref{nfromc0}, only depends on $c_0$. Hence, one could think about reducing the cosmic neutrino number density by reducing $c_0$, which could explain why the cosmological sum of neutrino masses has not been seen and therefore ease the tension with oscillation experiments. However, as we see in Fig.~\ref{fig:CnuB_constraints_compared}, constraints from BBN abundances do not allow $c_0$ to shrink significantly below its standard value, so they disfavor such a solution to the neutrino mass tension. Nevertheless, a varying $c_0$ that took its standard value at BBN times and decreased later---as would be the case, for example, in neutrino decay models (see~\cite{2024JHEP...09..097C_negativenumass} and references therein)---could be a viable solution, but such discussions are outside the scope of this paper.

\subsection{CMB distortions}
\subsubsection{Standard model distortion}

Even though the CMB spectrum is very close to a blackbody with temperature $T_\mathrm{CMB}$, it can still exhibit spectral distortions, defined as the difference between the real CMB spectrum and a blackbody spectrum at $T_\mathrm{CMB}$. The standard cosmological model ($\Lambda\mathrm{CDM}$) predicts such distortions to be nonzero, albeit very small \cite{Sunyaev:2013aoa, Chluba:2011hw, Chluba:2019kpb, Stebbins:2007ve}, because of the differential cooling of the relativistic photons by the non-relativistic baryons during cosmological expansion~\cite{Pitrou:2019hqg}. Besides, several non-standard cosmological models predict different distortions \cite{Chluba:2011hw, Chluba:2019kpb, Chluba:2020oip, Li:2025clq}. The current constraints on spectral distortions set by COBE / FIRAS \cite{Fixsen:1996nj} are upper bounds, but an actual observation of these distortions could be made by future experiments such as PIXIE~\cite{Kogut:2024vbi, Kogut:2019vqh, 2011JCAP...07..025K} and others~\cite{Sabyr:2024lgg, Maffei:2021xur, 2024JInst..19P2040A, 2021arXiv211012254M, PRISM:2013fvg}.

CMB distortions are usually characterized in terms of three theoretically well-motivated components: temperature difference, $\mu$, and $y$ distortions. However, these components do not capture the full distortions in most models, and a residual component is expected to exist \cite{Chluba:2013pya, Chluba:2013vsa, Khatri:2012tw}. This residual distortion is often characterized via a principal component analysis \cite{Chluba:2013pya, Lucca:2019rxf}, but this analysis is detector-dependent and often truncated after a couple of terms. Here, we use our OPE framework as an alternative model- and detector-independent approach to characterize the full distortions.

Since CMB distortions are so small, it is numerically challenging to calculate directly the OPE of the full spectrum. We therefore split the distribution as in Eq.~\eqref{eq:distorted_spectrum}, taking as a reference temperature the CMB temperature $\bref = 1/\kB T_\mathrm{CMB}$. In the following, we focus on the decomposition of the distortions alone, which are orders of magnitude smaller than the blackbody part. Note that the coefficients describing the full spectrum are the same, except for $c_0$ which is increased by $1/A_{00}$.

The left panel of Fig.~\ref{fig:standard_distortions_poly_decomp} shows the polynomial decomposition of the standard spectral distortions. The spectrum of spectral distortions was computed with CLASS~\cite{2011arXiv1104.2932L_Class1, 2011JCAP...07..034B_Class2} — see the details of the origin and treatment of CMB distortions in CLASS in~\cite{Lucca:2019rxf}. Even truncated to few terms, the OPE provides an accurate reconstruction of the distortion. The main noticeable problem is an error in the negative peak at very low frequency, which does not improve if we add more terms, and is likely due to the limited energy resolution of the spectral distortion spectrum provided by CLASS.

\subsubsection{Energy injection in the plasma}

Additional CMB distortions from possible energy injections have been characterized with Green functions \cite{Chluba:2013vsa}, defined as
\begin{equation}
    \Delta I_\nu (\nu) = \int G_\mathrm{th}(\nu,z)\frac{\mathrm{d}(Q/\rho_\gamma)}{\mathrm{d}z} \mathrm{d}z\,,
    \label{Greens_function_definition}
\end{equation}
where $\Delta I_\nu (\nu)$ is the observed intensity of the CMB spectral distortion at redshift $z=0$, $G_\mathrm{th}(\nu,z)$ is the Green's function, and $\mathrm{d}(Q/\rho_\gamma)/\mathrm{d}z$ is the so-called ``energy injection history," which encodes the amount of energy that is injected into the plasma at each redshift. This expression involving a Green's function is particularly useful since, if one assumes a fixed background cosmology, the Green's function is fixed and only needs to be computed once. It can then be used to compute the distortions associated with any standard or nonstandard energy injection. From Eq.~\eqref{Greens_function_definition}, we see that a physical way to interpret the Green's function is to state that $G_\mathrm{th}(\nu,z)$ equals the distortion caused by a Dirac-delta-shaped energy injection at redshift $z$.

As the temperature of the Universe decreases and interactions in the plasma change, different regimes appear for the injected spectral distortions, and these regimes are reflected in the redshift dependence of the Green's function. The three main regimes for the distortions are: temperature shifts ($z \gtrsim 2 \cdot 10^6$), $\mu$ distortions ($3 \cdot 10^5 \lesssim z \lesssim 2 \cdot 10^6$) and $y$ distortions ($z \lesssim 10^4$). Between the latter two redshift ranges, there is an intermediate regime where distortions do not have a known analytical form, nor are they a linear combination of $\mu$ and $y$ distortions. As a consequence, they need to be characterized numerically, which is usually done in a detector-dependent way via a principal component analysis~\cite{Chluba:2013pya, Lucca:2019rxf}.

Here, we instead employ our OPE framework as an alternative detector-independent method to characterize these intermediate distortions. To that end, we take the Green's function from CLASS~\cite{2011arXiv1104.2932L_Class1, 2011JCAP...07..034B_Class2, Lucca:2019rxf} and evaluate it at a redshift $z = 4 \cdot 10^4$, where intermediate distortions are expected to be the most important~\cite{Chluba:2013pya}. The actual spectrum, a $\mu$ + $y$ fit and several truncations of the OPE are shown on the right panel of Fig.~\ref{fig:standard_distortions_poly_decomp}. We see that a few polynomials can accurately characterize these distortions, without requiring to invoke a detector-dependent principal component analysis on top of the $\mu$+$y$ fit.

\begin{figure}[!ht]
    \centering
    \includegraphics[width=\columnwidth]{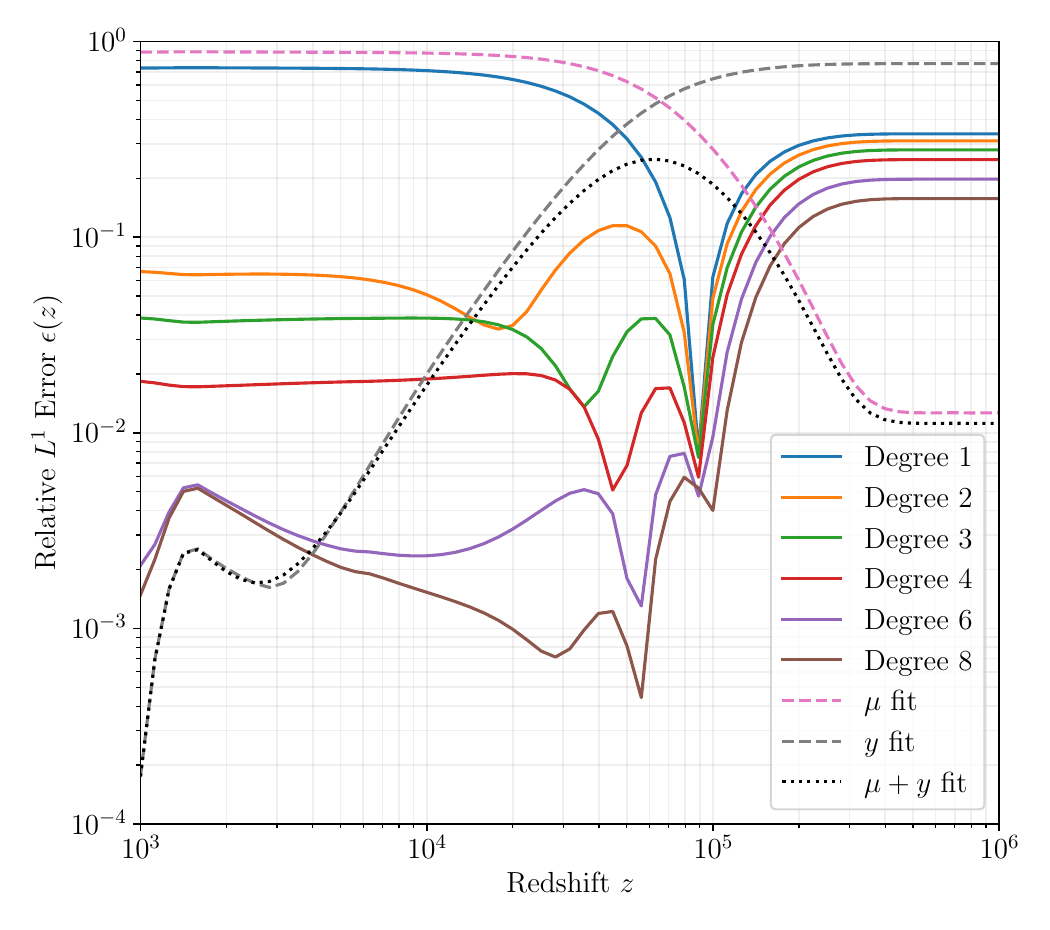}
    \caption{Relative $L^1$ error, defined in~\eqref{L1_relerror_def}, of the approximation to CMB distortions caused by a $\delta$-shaped energy injection at redshift $z$. Only deviations from a blackbody are shown here, and temperature difference distortions are not included. We show polynomial approximations of various degrees, as well as fits to $\mu$-distortions, $y$-distortions and a combination of both. For the polynomials and for the error calculation, we employ a cutoff $\nu_{\min} = 10 \, \mathrm{GHz}$. As expected, a $y$-distortion is the best fit at low redshifts, whereas a $\mu$-distortion works better at high redshifts. The $\mu$-distortion low-energy behavior is difficult to capture for the polynomials, which converge very slowly in this regime, as anticipated in Fig.~\ref{fig:muF_dist_polydecomp}, right panel. On the other hand, the OPE method is ideal to describe the CMB distortions in the intermediate redshift regime.}
    \label{fig:cmbdis_redshift_errors}
\end{figure}

We can carry out the same analysis for any energy-injection redshift. To characterize the error of an approximation to a function $\mathcal{S}$ with a single number, we employ the $L^1$ distance, and define the $L^1$ relative error as
\begin{equation}
    \epsilon(z) = \frac{\int | \mathcal{S}_\mathrm{exact}(x, z) - \mathcal{S}_{\mathrm{approx}}(x, z) | \mathrm{d}x}{\int | \mathcal{S}_\mathrm{exact}(x, z) | \mathrm{d}x} \, .
    \label{L1_relerror_def}
\end{equation}
In Fig.~\ref{fig:cmbdis_redshift_errors}, we show how the $L^1$ relative error of the different approximations evolves with redshift. We show different truncations of the OPE, as well as fits to $\mu$ distortions, $y$ distortions, and combinations thereof. For the error calculation of the OPE method, we employ a cutoff $\nu_{\min} = 10 \, \mathrm{GHz}$. This avoids a bad numerical behavior caused by the lowest-energy part of the spectrum which is anyway irrelevant in practice since the CMB is not expected to be observable below that frequency for current or future experiments.\footnote{For comparison, the lowest energy at which PIXIE expects to observe the CMB is $28 \, \mathrm{GHz}$~\cite{Kogut:2024vbi}. Only TMS expects to see the CMB at $\nu_{\min} = 10 \, \mathrm{GHz}$~\cite{2024JInst..19P2040A}.} As expected, $\mu$ (resp. $y$) distortions describe very well the high (resp. low) redshift regime, whereas intermediate redshift energy injections cannot be described by either of them, nor by their combination. For these intermediate redshifts, the OPE is ideal. In the $y$-distortion regime at low redshifts, the OPE performs very well, consistent with the pure $y$-distortion case studied in Sec.~\ref{subsec:ydistortion}. The comparatively poor convergence of the OPE error for high redshifts is due to the larger difficulty of the OPE method to describe $\mu$-distortions for bosons, see Sec.~\ref{subsec:mudistortion}. However, one could parametrize distortions by a combination of a $\mu$-type spectrum and an OPE, which would optimize the accuracy of the description without the need to use a principal component analysis.

\begin{figure}[!ht]
    \centering
    \includegraphics[width=\columnwidth]{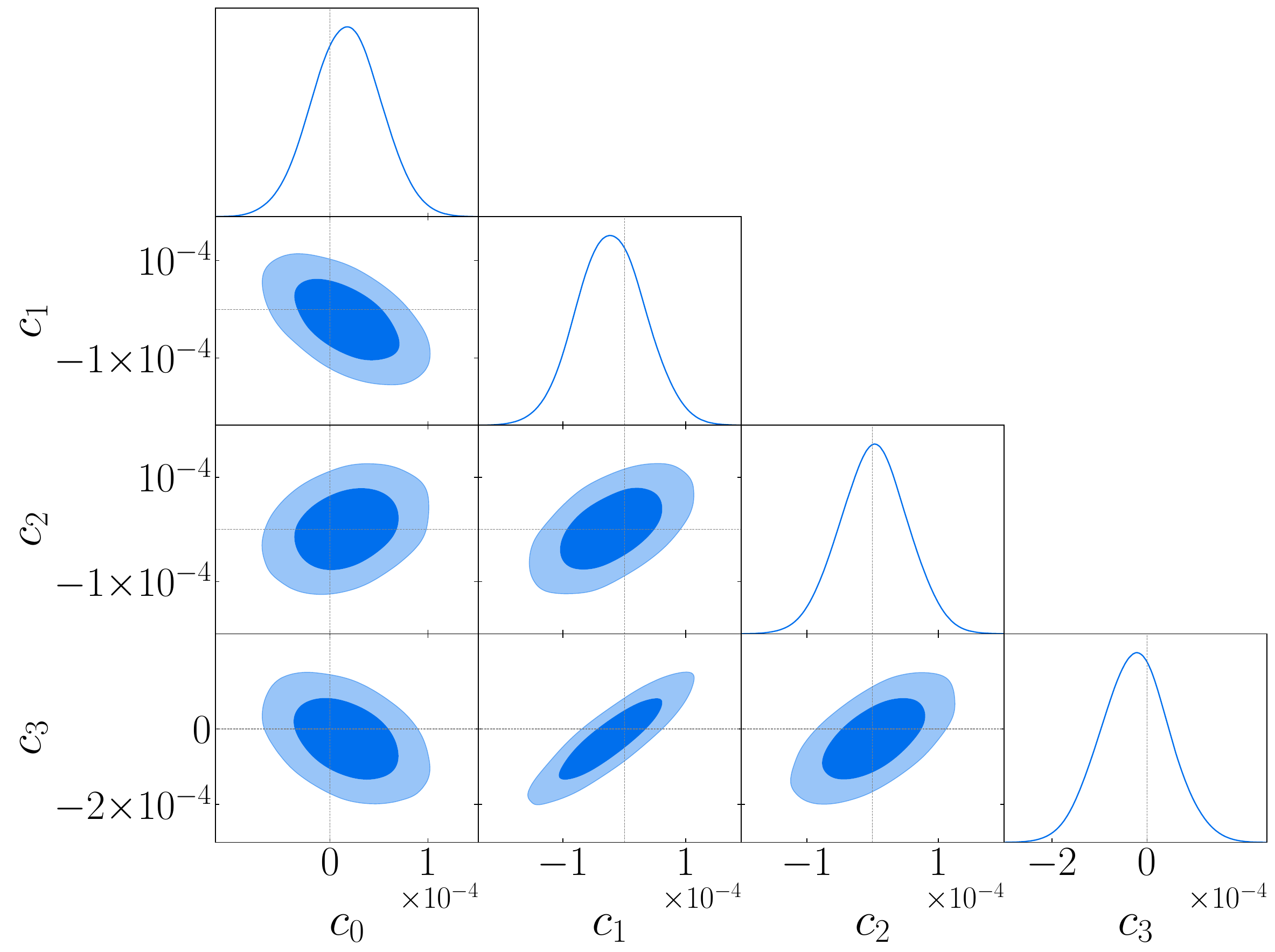}
    \caption{Constraints from COBE / FIRAS data \cite{Fixsen:1996nj} on the coefficients of the OPE of the CMB distortions, assuming the decomposition can be truncated to order 3. All coefficients are compatible with zero, meaning that no distortion is detected and the measured spectrum is fully compatible with a pure blackbody.}
    \label{fig:FIRAS_c3_constraints}
\end{figure}

Similarly to what we did for the C$\nu$B in Sec.~\ref{subsec:CnuB}, we constrain the possible CMB spectral distortions using our OPE framework, truncating the expansion to its first four terms. Except for the truncation error, which should be small for all reasonable spectra, this serves as a way to obtain model-independent bounds on the spectrum of CMB distortions. We conducted an MCMC analysis with the code Cobaya~\cite{Torrado:2020dgo}, that compares the spectrum associated with each set of coefficients with the data available from COBE / FIRAS \cite{Fixsen:1996nj}. We present the results in Fig.~\ref{fig:FIRAS_c3_constraints}. We see that all coefficients are compatible with zero, in accordance with the fact that no CMB spectral distortion was detected by FIRAS.

\section{Extension to massive case}
\label{Massive_Case}

So far we have assumed that particles are massless, or ultrarelativistic. For nonrelativistic massive particles, the previous framework is not particularly suitable. Let us introduce the classical kinetic energy $E_c$ by
\begin{equation}
E\simeq m c^2 + E_c\,,\quad E_c \equiv \frac{p^2}{2 m}\,,
\end{equation}
such that $\dd E_c = (p/m)\dd p =\sqrt{(2 E_c/m)}\dd p$. The phase space element $p^2 \dd p$ therefore becomes $\sqrt{2 m^3 E_c} \dd E_c$. In order to build a suitable basis for nonrelativistic particles, we must however distinguish between the degenerate case and the classical case so as to choose the appropriate weight in~\eqref{ScalarProduct_def}. 

For the case of degenerate fermions\footnote{The more complicated case of degenerate massive bosons, which form a Bose-Einstein condensate, is not discussed here.}, we can expand around a thermal spectrum with $\mu = m c^2$, so that $E-\mu=E_c$. By using $x\equiv E_c/\kB T$ as a variable instead of $E/\kB T$ we see that the natural weight is $w(x) = \sqrt{x}/({\rm e}^x\pm 1)$. The orthonormal polynomials $Q_i(x)$ (and thus the coefficients $A^Q_{ip}$ defined by $Q_i(x) \equiv \sum_{p=0}^i A^Q_{ip}x^p$) associated with this weight are built as the $P_i(x)$, but with the Hankel matrix components $H_{pq} = I_{p+q+1/2}$ instead of~\eqref{HankelValues}. The distribution function is then expanded similarly to~\eqref{Spectrum_decomposition} for fermions as
\begin{equation}
\begin{aligned}
     \sqrt{m^3 E_c} f(E_c) &=\frac{2\pi^2\hbar^3}{\sqrt{2}g_s} n(E_c) \\
     &=\frac{\sqrt{m^3 E_c}}{e^{\beta_\mathrm{ref}E_c} + 1} \sum_{i=0}^\infty c_i \, Q_i\left(\beta_\mathrm{ref}E_c\right)\,,
\end{aligned}
    \label{Spectrum_decomposition_massive}
\end{equation}
where the total number density is given by $n = \int_0^\infty n(E_c) \dd E_c$. The coefficients are extracted similarly to~\eqref{coeffs_calc}, namely,
\begin{align}
    c_i &= \beta_\mathrm{ref}^{3/2} \int_0^\infty \sqrt{E_c} f(E_c) Q_i\left(\beta_\mathrm{ref}E_c\right)\mathrm{d}E_c\nonumber\\
    &= \sum_{p=0}^i A^Q_{ip} \beta_\mathrm{ref}^{3/2+p}f_{p+1/2}\,,
    \label{coeffs_calc_massive}
\end{align}
where the classical energy moments are $f_n\equiv \int_0^\infty E_c^n f(E_c) \dd E_c$.

In the classical case, the suitable approximation is a thermal spectrum with $\mu \ll m c^2$ such that $f(E) \ll 1$. This reference distribution function is then well approximated by a Maxwell-Boltzmann distribution, that is 
\begin{equation}
\frac{1}{\exp\left(\dfrac{E-\mu}{\kB T}\right) \pm 1} \simeq \exp\left(\dfrac{\mu-m c^2-E_c}{\kB T}\right) \, .
\end{equation}
Therefore the natural weight to build a set of orthonormal polynomials is $w(x) = \sqrt{x}{\rm e}^{-x}$ and the Hankel matrix components are $H_{pq} = \Gamma(p+q+3/2)$. The orthonormal basis is actually known in this case and is given by the set of normalized associated Laguerre polynomials 
\begin{equation}\label{LaguerreMB}
Q^{\rm MB}_n(x) \equiv (-1)^n L^{(1/2)}_n(x)\sqrt{\frac{n!}{\Gamma(n+3/2)}}\,\,.
\end{equation}
The polynomial components defined by $Q^{\rm MB}_n(x) = \sum_{p=0}^n A^{\rm MB}_{np} x^p$ are
\begin{equation}
A^{\rm MB}_{np}= (-1)^{p+n}\frac{\sqrt{\Gamma(n+3/2) n!}}{(n-p)! \, p! \, \Gamma(p+3/2)}\,.
\end{equation}
The decomposition of the classical energy distribution function is similar to~\eqref{Spectrum_decomposition_massive} without the $+ 1$ in the denominator, and with $Q_i \to Q^{\rm MB}_i$, and the coefficient extraction is given by~\eqref{coeffs_calc_massive} with $A^Q_{ip} \to A^{\rm MB}_{ip}$.

\section{Conclusion}\label{Conclusion}

\begin{figure*}[!ht]
    \includegraphics[width=0.8\textwidth]{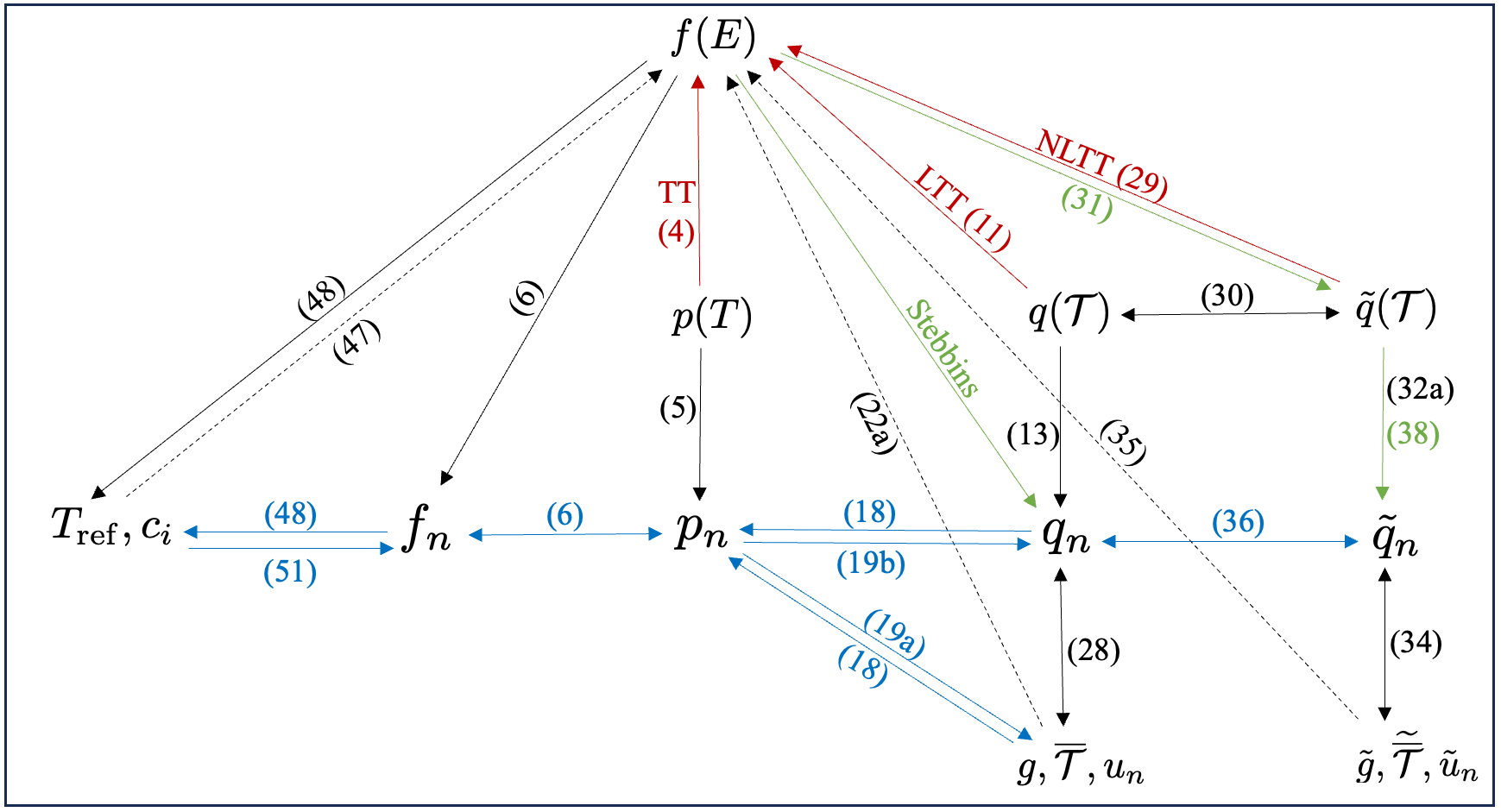}
    \caption{Summary of the constitutive relations in the different formalisms used to describe spectral distortions. The various definitions of temperature transforms are depicted with red arrows. Calculation of moments or extraction of coefficients from the spectrum are depicted with black arrows, and relations between moments are depicted with blue arrows. The practical methods to obtain the temperature transform quantities from the spectrum are shown with green arrows. They can either be Stebbins' method which estimates directly the LTT moments $q_n$, or our Fourier transform method which estimates first the NLTT $\tilde{q}({\cal T})$ and then extracts its moments via the moment-generating function. Finally, the approximate spectrum reconstructions are depicted with dashed arrows.}
    \label{fig:SummaryChart}
\end{figure*}

Distortions in the distribution function of a particle species is the signature of out-of-equilibrium processes. Theoretical predictions and experimental constraints are easily compared if distortion shapes can be approximately described by a few parameters. A generic idea by Stebbins~\cite{Stebbins:2007ve} consists in describing a distorted spectrum as a sum of thermal spectra, the associated distribution of temperatures being called the ``temperature transform." It is then possible to describe the spectrum from the moments of the temperature transform. The log-temperature transform summarized in Sec.~\ref{SecLTT} is particularly suited when gravitational effects redshift the energies, since the associated central moments remain invariant. However, it is numerically challenging to extract the moments from a given distribution, rendering these methods hardly usable in practice. In Sec.~\ref{SecNLTT}, we generalized those existing temperature transforms by working at the level of the number density distribution $E^2 f(E)$, which removes numerical instabilities at low energies. With this ``number density log-temperature transform" (NLTT), we can use Fourier transform techniques to easily go from the distribution to the NLTT moments, and vice versa. We summarize in Fig.~\ref{fig:SummaryChart} the relations between the distribution function and the various temperature transform moments.

An independent approach, which we found to be more stable numerically and applicable to more cases, consists in expanding the number density distribution on a set of orthonormal polynomials, which are adapted to the particle statistics (fermions/bosons), see Sec.~\ref{SecPE}. Increasing order coefficients of this Orthonormal Polynomial Expansion (OPE) describe increasing order energy moments of the distributions, such that truncating the OPE is a natural way to approximately describe the spectrum. We verified (Sec.~\ref{SecIdeal}) that the OPE method with a few coefficients is sufficient to capture the main distortion shapes, whereas the NLTT method fails for $\mu$-type distortions. 
We thus advocate for the use of the OPE method as a model-independent, instrument-independent parametrization of distortions that can be implemented in many different setups. As a proof of concept, we described in Sec.~\ref{SecApplications} the C$\nu$B and the CMB in the OPE framework, which works particularly well given the small size of distortions involved. Furthermore, this formalism is very well suited to constrain nonstandard scenarios which would result in large distortions.

For neutrinos, using the BBN code \primat, we computed the primordial abundances of helium-4 and deuterium, scanning through the parameter space of the $\{c_i\}$ coefficients of the OPE, see Fig.~\ref{fig:output_PRIMAT}. Using the most recent BBN spectroscopic measurements and CMB constraints of the effective number of relativistic species, we obtained in Fig.~\ref{fig:CnuB_constraints_compared} the current constraints on spectral distortions of the C$\nu$B, expressed in the model-independent language of the OPE.

For photons, we showed that the OPE can describe the range of distortions that result from a redshift-localized energy injection, as summarized in Fig.~\ref{fig:cmbdis_redshift_errors}. In particular, for an energy injection in the ``intermediate" redshift regime ($2 \times 10^4 < z < 3 \times 10^5$), where the resulting distortions are neither of the $\mu$ or $y$ type, our approach is very competitive. We note that the standard description of CMB distortions in CLASS in this regime involves a principal component analysis, which suffers from being instrument-dependent. As shown in the simple case of a pure bosonic $\mu$ distortion in Sec.~\ref{subsec:mudistortion}, the OPE method encounters the most difficulties in the $\mu$ distortion regime, but the error is still reduced by increasing the degree of truncation, and the use of a spectrum with chemical potential as a reference would be a natural extension of the method. We also obtained the constraints on the OPE coefficients using the data from FIRAS, which show as expected that the spectrum is fully compatible with a blackbody. Future experiments looking specifically at spectral distortions of the CMB (see e.g.,~\cite{Kogut:2024vbi, Kogut:2019vqh, 2011JCAP...07..025K,Sabyr:2024lgg, Maffei:2021xur, 2024JInst..19P2040A, 2021arXiv211012254M, PRISM:2013fvg}) will strongly improve these constraints, and the OPE provides a natural framework to express them. The OPE approach is easily adapted for nonrelativistic matter, as we detailed in Sec.~\ref{Massive_Case}.

Although we have focused on cosmological applications, we emphasize that the improved NLTT or the OPE approach are far more general. They could be applied across a wide range of contexts, from astrophysical signals (such as the diffuse supernova neutrino background spectrum~\cite{Beacom:2010kk}) to condensed matter systems (e.g., the electron blackbody spectrum of vertically aligned carbon nanotube arrays~\cite{2023PNAS..12009670Z}). Overall, these methods provide a powerful and flexible framework for describing generic spectra in cosmology and beyond.

\begin{acknowledgments}

G.B. and H.S. are supported by the Spanish grants  CIPROM/2021/054 (Generalitat Valenciana),  PID2023-151418NB-I00 funded by MCIU/AEI/10.13039/501100011033, and by the European ITN project HIDDeN (H2020-MSCA-ITN-2019/860881-HIDDeN). H.S. is also supported by the
grant FPU23/00257, MCIU., and acknowledges the support of the European Consortium for Astroparticle Theory in the form of an Exchange Travel Grant. J.F. acknowledges support from the Severo Ochoa Excellence Grant CEX2023-001292-S funded by MICIU/AEI/10.13039/501100011033, and from the Network for Neutrinos, Nuclear Astrophysics and Symmetries (N3AS), through the National Science Foundation Physics Frontier Center award No.~PHY-2020275. J.F. and H.S. thank the Institut d'Astrophysique de Paris for its hospitality close to the completion of this work. 

\end{acknowledgments}

\appendix
\section{Determination of polynomial coefficients}
\label{ap:PolynomialCoefficients}

The orthonormal polynomials that serve as a basis for the expansion we develop in Sec.~\ref{SecPE} are of the form $P_i(x) = \sum_{p=0}^i A_{ip} x^p$. By applying the Gram-Schmidt orthogonalization procedure, these coefficients can be determined up to a desired $n_\text{max}$. Their expressions are analytic, involving Riemann $\zeta$ functions. For instance 
\begin{equation}\label{ValueA00}
A_{00}=\frac{1}{\sqrt{I_2}} = 
\begin{cases} 
\sqrt{2/[3\zeta(3)]}\quad&\text{[fermions]}\,,\\
1/\sqrt{2\zeta(3)}\quad&\text{[bosons]}\,.
\end{cases}
\end{equation}
The next coefficients are
\begin{equation}\label{ValueA10A11}
A_{10}=\frac{-I_3}{I_2\sqrt{I_4-I_3^2/I_2}}\,, \quad
A_{11}=\frac{1}{\sqrt{I_4-I_3^2/I_2}}\,,
\end{equation}
whose explicit expressions require to replace the integrals~\eqref{DefIn}.
However, the complexity of these analytic expressions increases dramatically with $n$, such that we report below numerical values. The first coefficients up to $n=4$ are

\begin{widetext}
\begin{align}
\bm{A} &\supset \left(
\begin{array}{ccccc}
 0.744718 & 0 & 0 & 0 & 0 \\
 -1.35311 & 0.429373 & 0 & 0 & 0 \\
 1.97441 & -1.2475 & 0.151237 & 0 & 0 \\
 -2.6059 & 2.46541 & -0.596317 & 0.0389148 & 0 \\
 3.24492 & -4.09116 & 1.48209 & -0.193148 & 0.00792035 \\
\end{array}
\right) &&\text{[fermions]} \, , \\ \nonumber \\
\bm{A} &\supset \left(
\begin{array}{ccccc}
 0.644945 & 0 & 0 & 0 & 0 \\
 -0.996685 & 0.368982 & 0 & 0 & 0 \\
 1.28254 & -0.975145 & 0.130561 & 0 & 0 \\
 -1.53019 & 1.78013 & -0.482551 & 0.0338102 & 0 \\
 1.7519 & -2.76026 & 1.13368 & -0.159954 & 0.00692472 \\
\end{array}
\right)  &&\text{[bosons]} \, .
\end{align}
\end{widetext}

\section{Recursive construction of the orthonormal polynomials}
\label{app:recursion}

The orthonormal polynomials \( P_n(x) \) used throughout this work are defined with respect to the scalar product~\eqref{ScalarProduct_def}, and satisfy~\eqref{Pi_orthogonality}.
These polynomials satisfy the standard three-term recurrence relation
\begin{equation}
    \beta_{n+1} P_{n+1}(x) = (x - \alpha_n) P_n(x) - \beta_n P_{n-1}(x)\,,
    \label{eq:threetermrecursion}
\end{equation}
with initial conditions
\begin{equation}
    P_{-1}(x) = 0\,, \qquad P_0(x) = \frac{1}{\sqrt{I_2}}\,.
\end{equation}
The recurrence coefficients are
\begin{align}
    \alpha_n = \Braket{xP_n | P_n} &= \int_0^\infty w(x)\, x\, P_n(x)^2\, \mathrm{d}x\,, \\
    \beta_n = \Braket{xP_n | P_{n-1}} &= \int_0^\infty w(x)\, x\, P_n(x)\, P_{n-1}(x)\, \mathrm{d}x\,.
\end{align}

Using~\eqref{eq:threetermrecursion}, one can compute $\beta_{n+1} P_{n+1}(x)$ from the previous polynomials and their coefficients. From there, one can get $\beta_{n+1}$ and $P_{n+1}(x)$ by normalizing $\beta_{n+1} P_{n+1}(x)$.\footnote{When normalizing, one can freely choose the sign of $\beta_{n+1}$, which in turn affects the sign of $P_{n+1}$. This is related to the freedom of choice of the sign of the orthonormal basis vectors. We choose $\beta_{n+1}$ to be positive. By looking at leading coefficients (note that the leading term of $\beta_{n+1} P_{n+1}(x)$ only gets a contribution from the leading term of $xP_n(x)$), one can see that this is equivalent to choosing our orthonormal polynomials such that the leading coefficient is always positive.} Note that this already yields the $\beta_n$ for the next step, so it is not required to compute the $\beta_n$ coefficients from their definition.

For polynomials with a very high degree, directly calculating them with the recurrence relation on the grid of desired $x$ values is numerically more stable than calculating their coefficients first and calculating the polynomials from the coefficients later. Indeed, high-degree polynomials require precise cancellations between terms that will heavily amplify any error in the coefficients, even tiny errors due to machine precision.

\bibliography{Paper.bib}
\end{document}